\def\BA{\begin{array}}
\def\BEA{\begin{eqnarray}}
\def\BE{\begin{equation}}
\def\BI{\begin{enumerate}}
\def\BML{\begin{mathletters}}

\def\EA{\end{array}}
\def\EEA{\end{eqnarray}}
\def\EE{\end{equation}}
\def\EI{\end{enumerate}}
\def\EML{\end{mathletters}}
\def\Frac#1#2{{\displaystyle #1\over\displaystyle #2}}
\def\Ham{{\cal H}}
\def\Jb{{\bar J}}
\def\Jv{{\bf J}}
\def\Jbv{{\bf\Jb}}
\def\Lag{{\cal L}}
\def\La{\Lambda}
\def\L{\langle}
\def\NN{\nonumber}
\def\Ocdw{{\cal O}_{\rm CDW}}

\def\R{\rangle}

\def\Zn{{\bf Z}}
\def\a{\alpha}
\def\bb{{\bar b}}
\def\b{\beta}
\def\com#1{\lbrack #1\rbrack}
\def\db{\bar\partial}
\def\dd#1#2{{\dr #1\over \dr #2}}
\def\dk#1{{\dr #1\over 2\pi}}
\def\dg{\dagger}

\def\dn{\downarrow}
\def\dual{\vartheta}
\def\dr{{\rm d}}
\def\dx{\dr x\;}
\def\dy{\dr y\;}
\def\d{\partial}
\def\epsb{{\bar\eps}}
\def\eps{\varepsilon}
\def\etab{{\bar\eta}}
\def\e{\,{\rm e}}
\def\fant{\phantom{-}}
\def\frac#1#2{{#1\over #2}}
\def\ffrac#1#2{{\textstyle{#1\over #2}}}
\def\g{\gamma}
\def\hb{{\bar h}}
\def\hf{\frac12}
\def\hff{\ffrac12}

\def\la{\lambda}
\def\l{\left}
\def\no#1{:\kern-2pt #1\kern-2pt:}
\def\ntot{n_{\rm tot.}}
\def\om{\omega}
\def\phib{\skew5\bar\phi}
\def\psib{\skew5\bar\psi}
\def\p{\varphi}
\def\r{\right}
\def\sd{\sqrt{2\pi}}
\def\sk{\sqrt{4\pi}}
\def\s{\sigma}
\def\t{\theta}
\def\tr{{\rm tr}}
\def\up{\uparrow}
\def\zb{{\bar z}}
\def\text#1{\quad\hbox{#1}\quad}

\def\figEPS#1#2{
\epsfxsize=#2\medskip\centerline{\epsfbox{#1}}\medskip
}

\documentstyle[epsf]{article}
\input epsf.tex

\begin{document}

\bibliographystyle{prsty}

\title{An introduction to bosonization}
\author{David S\'en\'echal\\
\small\it Centre de recherches en physique du solide
and D\'epartement de physique\\
\small\it Universit\'e de Sherbrooke,
Sherbrooke (Qu\'ebec) Canada J1K 2R1
}
\date{August 1999, Report no CRPS-99-09 {\tt}}
\maketitle

\begin{abstract}
This is an expanded version of a lecture given at the {\it Workshop on Theoretical
Methods for Strongly Correlated Fermions}, held at the {\it Centre de Recherches
Math\'ematiques}, in Montr\'eal, from May 26 to May 30, 1999. After general
comments on the relevance of field theory to condensed matter systems, the
continuum description of interacting electrons in 1D is summarized. The
bosonization procedure is then introduced heuristically, but the precise
quantum equivalence between fermion and boson is also presented. Then the exact
solution of the Tomonaga-Luttinger model is carried out. Two other applications
of bosonization are then sketched. We end with a quick introduction to
non-Abelian bosonization.
\end{abstract}

\tableofcontents

\section{Quantum Field Theory in Condensed Matter}

Before embarking upon a technical description of what bosonization is and how
it helps understanding the behavior of interacting electrons in one dimension,
let us make some general comments on the usefulness of quantum field theory in
the study of condensed matter systems. First of all, what is a quantum field
theory? From the late 1920's to the 1980's, it has taken over half a century
before a satisfactory answer to this fundamental question creep into the
general folklore of theoretical physics, the most significant progress being
the introduction of the renormalization group (RG). A field theory is by
definition a physical model defined on the continuum as opposed to a lattice.
However, quantum fluctuations being at work on all length scales, a purely
continuum theory makes no sense, and a momentum cutoff $\La$ (or another
physically equivalent procedure) has to be introduced in order to enable
meaningful calculations. Such a {\it regularization} of the field theory
invariably introduces a length scale $\La^{-1}$, which is part of the
definition of the theory as one of its parameters, along with various coupling
constants, masses, and so on. A change in the cutoff $\La$ (through a trace
over the high-momentum degrees of freedom) is accompanied by a modification of
all other parameters of the theory. A field theory is then characterized not
by a set of fixed parameter values, but by a RG trajectory
in parameter space, which traces the changing parameters of the theory as the
cutoff is lowered.

Most important is the concept of {\it fixed point}, i.e., of a theory whose
parameters are the same whatever the value of the cutoff. Most notorious are
free particle theories (bosons or fermions), in which degrees of freedom
at different momentum scales are decoupled, so that a partial trace in a
momentum shell has no impact on the remaining degrees of freedom. Theories
close (in a perturbative sense) to fixed points see their parameters fall into
three categories: relevant, irrelevant and marginal. Relevant parameters grow
algebraically under renormalization, irrelevant parameters decrease
algebraically, whereas marginal parameters undergo logarithmic variations.
It used to be that only theories without irrelevant
parameters were thought to make sense -- and were termed {\it renormalizable} --
for the following reason: If the momentum cutoff is taken to infinity, i.e.,
if the starting point of the renormalization procedure is at arbitrarily high
energy, then an arbitrary number of irrelevant parameters can be added to the
theory without measurable effect on the low-energy properties determined from
experiments. Thus, the theory has no predictive power on its irrelevant
parameters and the latter were excluded for this reason. In a more modern view,
the cutoff $\La$ does not have to be taken to infinity, but has some natural
value $\La_0$, determined by a more microscopic theory (maybe even another
field theory) which eventually supersedes the field theory considered at length
scales smaller than $\La^{-1}$. In the physics of fundamental interactions, the
standard model may then be considered an effective field theory, superseded by
something like string theory at the Planck scale ($\sim 10^{-33}$cm). Likewise,
effective theories for the strong interactions are superseded by QCD at the QCD
length scale ($\sim 10^{-13}$cm). In condensed matter physics, the natural
cutoff is the lattice spacing ($\sim 10^{-8}$cm).

In practice, field theories should not be pushed too close to their natural
cutoff $\La_0$. It is expected that a large number (if not an infinity) of
irrelevant couplings of order unity exist at that scale, and the theory then
loses all predictive power. The general practice is to ignore irrelevant
couplings altogether, and this is credible only well below the natural
cutoff.\footnote{In some applications, leading irrelevant parameters are
physically important and must be kept.} The price to pay for this reduction in
parameters is that the finite number of marginal or relevant parameters
remaining cannot be quantitatively determined from the underlying microscopic
theory (i.e. the lattice model). However, the predictions of the field theory
can (in principle) be compared with experiments and the parameters of the
theory be inferred. After all, this is what the standard model of elementary
particles is about. But even in cases where a quantitative determination of all
the parameters of the theory is not possible, some universal (i.e., unaffected
by the details of the microscopic theory) predictions are possible, and are the
main target of fields theory of condensed matter systems.

The connection between condensed matter systems and relativistic field theories
is particularly fruitful in one dimension, in good part because the finite
electron density may be swept under the rug (or rather, the Fermi sea), leaving
low-energy excitations enjoying a Lorentz-like invariance in the absence of
interactions (relativistic field theories are generally applied at zero
density). This is not true in higher dimensions, where the shape of the Fermi
surface makes the connection with relativistic field theories more difficult.
The development of bosonization was pursued in parallel by condensed matter and
particle physicists. The former had the one-dimensional electron gas in mind,
whereas the latter were first interested in a low-dimensional toy model for
strong interactions, and later by the applications of bosonization to string
theory. Ref.~\cite{Stone94} reproduces many important papers in the development
of bosonization, along with a summary of the method (in a language more
appropriate for particle theorists) and a few historical remarks.

\section{A word on conformal symmetry}

\subsection{Scale and conformal invariance}

Let us briefly consider the action of the renormalization group (RG) on
fermions. The low-energy limit is defined around the Fermi surface. The
RG action consists in progressively tracing over degrees of freedom far
from the Fermi surface, towards the Fermi surface. In one dimension, this
process may be understood better by considering Fig.~\ref{fermiFIG} below.
After an RG step wherein the degrees of freedom located in the momentum shell
$[\La-\dr\La,\La]$ have been traced over, one may choose to perform an
infinitesimal momentum scale transformation $k\to(1+\dr\La/\La)k$ which brings
the cutoff back to its initial value $\La$ before this RG step. This rescaling
is necessary if one wants to compare the new coupling constants with the old
ones (i.e. if one wants to define a RG trajectory) since we are then comparing
theories with identical cutoffs $\La$. In a (nearly) Lorentz invariant theory,
it is natural to perform this scale transformation both in  momentum and energy
(or space and time).

In the context of one-dimensional electron systems, Lorentz invariance is an
emerging symmetry, valid in the low-energy limit, when the linear
approximation to the dispersion relation is acceptable. The Fermi velocity
$v$ then plays the role of the velocity of light: it is the only
characteristic velocity of the system. Interactions may violate
Lorentz invariance and cause the appearance of two characteristic velocities
$v_s$ and $v_c$ (spin-charge separation). However, again in the low-energy
limit, the system may then separate into disjoint sectors (spin and charge),
each benefiting from Lorentz invariance, albeit with different ``velocities of
light''. In the imaginary-time formalism, Lorentz invariance becomes
ordinary spatial-rotation invariance.

A field theory with rotation, translation and scale invariance is said to have
{\it conformal invariance}. Fixed point theories,
such as the free boson or the free fermion theories (and many interacting
theories) are thus {\it conformal field theories}. A decent introduction to
conformal field theory is beyond the scope of such a short tutorial. The
subject is quite vast and has ramifications not only in the field of strongly
correlated electrons, but also in the statistical physics of two-dimensional
systems, in string theory and in mathematical physics in general. We will
simply mention minimal implications of conformal symmetry in bosonization.
Of course, departures from fixed points break this symmetry partially, but
their effect at weak coupling may be studied in the conformal symmetry
framework with profit.

\subsection{Conformal transformations}

By definition, a conformal transformation is a mapping of space-time
unto itself that is locally equivalent to a rotation and a dilation. In two
dimensions, and in terms of the complex coordinate $z$, it can be shown that the
only such transformations have the form
\BE
z\to {az+b\over cz+d}\qquad ad-bc=1
\EE
where the parameters $a,b,c,d$ are complex numbers. In dimension $d$, the
number of independent parameters of such transformations is $(d+1)(d+2)/2$,
and $d=2$ is no exception. However, the case of two dimensions is very
special, because of the possibility of {\it local} conformal transformations,
which have all the characters of proper conformal transformations, except that
they are not one-to-one (they do not map the whole complex plane onto itself).
Any analytic mapping $z\to w$ on the complex plane is locally conformal, as we
known from elementary complex analysis. Indeed, the local line element on the
complex plane transforms as
\BE
\dr w = \l(\dd wz\r) \dr z
\EE
The modulus of the derivative embodies a local dilation, and its phase a local
rotation. This distinctive feature of conformal symmetry  in two-dimensional
space-time is what allows a complete solution of conformal field theories, even
in circumstances that apparently break scale invariance. For instance, the
entire complex plane (the space-time used with imaginary time at zero
temperature) may be mapped onto a cylinder of circumference $L$ via the complex
mapping $z=\e^{2\pi w/L}$. This allows for the calculation of correlation
functions in a system with a macroscopic length scale (a finite size $L$ at zero
temperature, or a finite-temperature $\b=L/v$ in an infinite system) from the
known solution in a scale-invariant situation. Mappings may also be performed
from the upper half-plane unto finite, open regions; such mappings are
particularly useful when studying boundary or impurity problems. On the formal
side, this feature of conformal field theory makes the use of complex
coordinates very convenient: In terms of $x$ and imaginary time $\tau=it$,
these coordinates may be defined as
\BE\label{zzb}
\BA{rll}
z &= -i(x-vt)  &= v\tau-ix \\
\zb &= \fant i(x+vt)  &= v\tau+ix 
\EA\EE
with the following correspondence of derivatives:
\BEA\label{zzb2}
\d_z &=& -{i\over2}\l( {1\over v}\d_t-\d_x\r)
\qquad \d_x = -i(\d_z-\d_\zb)\NN\\
\d_\zb &=& -{i\over2}\l( {1\over v}\d_t+\d_x\r) \qquad  \d_t =
iv(\d_z+\d_\zb) 
\EEA

A local operator $O(z,\zb)$ belonging to a conformal field theory is called
{\it primary} (more precisely, quasi-primary) if it scales well under
a conformal transformation, i.e., if the transformation
\BE
O'(\a z,\bar\a\zb) = \a^{-h}\bar\a^{-\hb}O(z,\zb)
\EE
is a symmetry of correlation functions. The constants $h$ and $\hb$ are a
fundamental property of the operator $O$ and are called the right- and
left-conformal dimensions, respectively. Under a plain dilation ($\a=\bar\a$),
the operator scales as $O'(\a z,\a\zb) = \a^{-\Delta}O(z,\zb)$, where
$\Delta=h+\hb$ is the ordinary scaling dimension of the operator. Under a
rotation (or Lorentz transformation), for which $\a=\e^{i\t}$ and
$\bar\a=\e^{-i\t}$, the operator transforms as $O'(\e^{i\t} z,\e^{-i\t}\zb) =
\e^{is\t}O(z,\zb)$, where $s=h-\hb$ is called the {\it conformal spin} of the
operator.

The scaling dimension appears directly in the two-point correlation function of
the operator, which is fixed by scaling arguments -- up to multiplicative
constant:
\BE\label{2point}
\L O(z,\zb)O^\dg(0,0)\R = {1\over z^{2h}}{1\over \zb^{2\hb}}
\EE

\subsection{Effect of perturbations}
Fixed-point physics helps understanding what happens in the vicinity of
the fixed point, in a perturbative sense.
Consider for instance the perturbed action
\BE
S = S_0 +  g\int \dr t\dr x\; O(x,t)
\EE
where $S_0$ is the fixed-point action and $O$ an operator of scaling dimension
$\Delta$. Let us perform an infinitesimal RG step, as described in the first
paragraph of this section, with a scale factor $\la=1+\dr \La/\La$ acting on
wavevectors; the inverse scaling factor acts on the coordinates, and if $x$ and
$t$ now represent the new, rescaled coordinates, with the same short-distance
cutoff as before the partial trace, then
\BEA
S' &=& S_0 + g\int \dr (\la t)\dr (\la x)\; O(\la x,\la t) \NN\\
&=& S_0 + g\la^{2-\Delta}\int \dr t\dr x\; O(x,t)
\EEA
We assumed that we are sufficiently close to the fixed point so that the effect
of the RG trace in the shell $\dr\La$ is precisely the scaling of $O$ with
exponent
$\Delta$. By integrating such infinitesimal scale transformations, we find that
the coupling constant $g$ scales as
\BE {g\over g_0} = \l({\La_0\over\La}\r)^{\Delta-2} =
\l({\xi_0\over\xi}\r)^{2-\Delta} 
\EE
where $\xi$ and $\xi_0$ are correlation lengths at the two cutoffs $\La$ and
$\La_0$. Thus, the perturbation is relevant if $\Delta<2$, irrelevant if
$\Delta>2$ and marginal if $\Delta=2$.

A relevant perturbation is typically the source of a gap in the
low-energy spectrum. In a Lorentz-invariant system, a finite correlation length
is associated with a mass gap $m\sim 1/\xi_0$. Let us suppose that the
correlation length is of the order of the lattice spacing ($\xi\sim 1$)
when the coupling constant $g$ is of order unity. The above scaling relation
then relates the mass $m$ to the bare (i.e. `microscopic') coupling constant:
\BE\label{gap}
m \sim g_0^{1/(2-\Delta)}
\EE
Given the warning issued in the introduction on the impossibility of knowing
the bare parameters of the theory with accuracy, we should add that `bare' may
mean here ``at a not-so-high energy scale where the coupling $g_0$ is known by
other means''. In fact, this scaling formula works surprisingly well even if
$g_0$ is taken as the bare coupling (at the natural cutoff $\La_0$), provided
$g_0$ is not too large.

Conformal field theory techniques may also be used to treat margi\-nal
perturbations, in performing the equivalent of one-loop calculations in
conventional perturbation theory, with the so-called operator product expansion.
This is explained in Appendix A and an example calculation is given (more
can be found in Refs \cite{Affleck89,Gogolin98,Allen97}).
However, perturbations with conformal spin $s\ne0$ are more subtle, and
typically lead to a change in $k_F$ or incommensurabilities.

\subsection{The central charge}
A conformal field theory is characterized by a number $c$ called the {\it
conformal charge}. This number is roughly a measure of the number of degrees of
freedom of the model considered. By convention, the free boson theory has
conformal charge
$c=1$. So does the free complex fermion. A set of $N$ independent free
bosons has central charge $c=N$. Conformal field theories with conformal charge
$c<1$ correspond to known critical statistical models, like the Ising model (known
since Onsager's solution to be equivalent to a free Majorana fermion), the
Potts model, etc. Models at $c=1$ practically all correspond to a free boson or
to slight modifications thereof called {\it orbifolds}, and have been
completely classified~\cite{Ginsparg88}. A conformal field theory of integer
central charge $c=N$ may have a representation in terms of $N$ free bosons,
although such representations are not known in general. Likewise, a conformal
field theory of half-integer central charge may have a representation in terms
of an integer number of real (i.e. Majorana) fermions.

\section{Interacting electrons in one dimension}

\subsection{Continuum fields and densities}

\begin{figure}
\figEPS{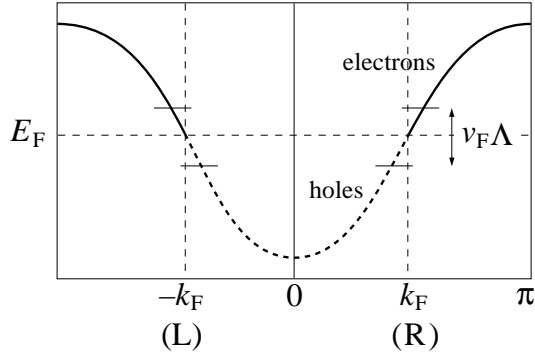}{7cm}
\caption{Typical tight-binding dispersion in 1D, illustrating left and
right Fermi points and the linear dispersion in the vicinity of those points.
}
\label{fermiFIG}
\end{figure}

Let us consider noninteracting electrons on a lattice and define the
corresponding (low-energy) field theory. This poses no problem, since the
different momentum scales of this free theory are decoupled. The microscopic
Hamiltonian takes the form
\BE
H_{\rm F} = \sum_k \eps(k) c^\dg(k) c(k)
\EE
where $c(k)$ is the electron annihilation operator at wavevector $k$ (we ignore
spin for the time being). The low-energy theory is defined in terms of creation
and annihilation operators in the vicinity of the Fermi points (cf.
Fig.~\ref{fermiFIG}) as follows:
\BEA
\a(k) &=& c(k_F+k)\cr
\a(-k) &=& c(-k_F-k)\cr
\b(k) &=& c^\dg(k_F-k)\cr
\b(-k) &=& c^\dg(-k_F+k)
\EEA
where $k$ is positive. The noninteracting Hamiltonian then takes the form
\BE\label{contHam1}
H_{\rm F} = \int\dk{k} v|k|\l\{ \a^\dg(k) \a(k) + \b^\dg(k)\b(k) \r\}
\EE
where $v$ is the Fermi velocity, the only remaining parameter from the
microscopic theory. The energy is now defined with respect to the ground state
and the momentum integration is carried between $-\La$ and $\La$.
The operators $\a(k)$ and $\b(k)$ respectively annihilate
electrons and holes around the right Fermi point ($k>0$) and the left Fermi
point ($k<0$). A momentum cutoff $\La$ is implied, which corresponds to an
energy cutoff $v\La$. Note that the continuum limit is in fact a low-energy
limit: Energy considerations determine around which wavevector ($\pm k_F$) one
should expand. In position space, this procedure amounts to introducing slow
fields $\psi$ and $\psib$ such that the annihilation operator at site $n$
is\footnote{We will generally use a bar ($\bar{\fant}$) to denote
left-moving operators, and the same symbol without the bar for right-moving
operators. A more common notation in condensed matter physics is the use of
subscripts $L$ and $R$. The (lighter) notation used here
stresses the analogy with complex coordinates.}
\BE
{c_x\over\sqrt{a}} = \psi(x)\e^{i k_F x} + \psib(x)\e^{-i k_F x}
\EE

The factor $\sqrt{a}$ is there to give the fields the proper delta-function
anticommutator, and reflects their (engineering) dimension:
\BEA\label{anticomm1}
\{\psi(x),\psi^\dg(x')\} &=& \delta(x-x') \cr
\{\psib(x),\psib^\dg(x')\} &=& \delta(x-x') \cr
\{\psi(x),\psib^\dg(x')\} &=& 0 
\EEA

\subsubsection*{Left-right separation}
The mode expansions of the continuum fields are
\BEA\label{fermimode1}
\psi(x) &=& \int_{k>0}\dk{k} \l[ \e^{ikx}\a(k) + \e^{-ikx}\b^\dg(k) \r] \cr
\bar\psi(x) &=& \int_{k<0}\dk{k} \l[ \e^{ikx}\a(k) + \e^{-ikx}\b^\dg(k) \r]
\EEA
The time dependence of $\a(k)$ and $\b(k)$ is obtained through multiplying by
the phase $\e^{-iv|k|t}$. In terms of the complex coordinates (\ref{zzb}),
the mode expansions for the time-dependent fields are then
\BEA
\psi(z) &=& \int_{k>0}\dk{k} \l[ \e^{-kz}\a(k) + \e^{kz}\b^\dg(k) \r] \NN\\
\psib(\zb) &=& \int_{k<0}\dk{k} \l[ \e^{k\zb}\a(k) + \e^{-k\zb}\b^\dg(k) \r]
\EEA
Thus the right-moving field $\psi$ depends solely on the right-moving
coordinate $z$, whereas the left-moving field $\psib$ depends solely on $\zb$. 

The mode expansion (\ref{fermimode1}) is misleading in one respect: it makes
believe that the positive-wavevector modes are contiguous to the
negative-wavevector modes in $k$-space, which is not the case (they are
separated by roughly $2k_F$). Thus right- and left-moving modes are clearly
well-separated. Left and right electrons have density fluctuations
\BE\label{e-currents}
J(x) = \psi^\dg(x)\psi(x) \qquad
\Jb(x) = \psib^\dg(x)\psib(x)
\EE
The total electronic density (with respect to the ground state)
being $\ntot=J+\Jb$.

\subsubsection*{Analogy with the Dirac equation}
It is a simple matter to express the continuum Hamiltonian (\ref{contHam1}) in
terms of the fields $\psi$ and $\psib$:
\BE\label{contHam2}
H_{\rm F} = -iv\int \dx \l[ \psi^\dg\d_x\psi  - \psib^\dg\d_x\psib\r]
\EE
This is exactly the Dirac Hamiltonian. Indeed, in (1+1) dimensions, the Dirac
matrices can be taken to be two-dimensional, for instance
\BE
\g^0 = \pmatrix{0&1\cr -1&0\cr}\qquad \g^1 = \pmatrix{0&1\cr 1&0\cr}
\EE
and the two-component Dirac spinor can be taken to be
\BE
\Psi = \pmatrix{\psi\cr\psib\cr}
\EE
The Dirac Lagrangian density -- with a mass term -- being
\BE
\Lag_D = i\bar\Psi(\g^0\d_t - v\g^1\d_x)\Psi - m\bar\Psi\Psi
\EE
The canonical Hamiltonian density derived from this Lagrangian is therefore
\BE\label{DiracH}
H_D =  -iv\l[ \psi^\dg\d_x\psi  - \psib^\dg\d_x\psib \r]
+ m(\psi^\dg\psib-\psib^\dg\psi)
\EE
which coincides with Eq.~(\ref{contHam2}) in the massless case.

Adding a mass term to the continuum Hamiltonian amounts to opening a gap
$2m$ at the Fermi level. On the other hand, adding a term like
\BE
\Ham' = \psi^\dg\psi + \psib^\dg\psib = J+\Jb
\EE
to the Hamiltonian would simply shift the chemical potential, i.e., modify
the value of $k_F$. This would of course require a redefinition of left and
right modes, but the end result would again be a Hamiltonian of the form
(\ref{contHam2}), with a slightly different velocity $v$.

In the language of Conformal Field Theory, the free fermion theory has
central charge $c=1$, and the conformal dimensions of the fermion fields
are
\BE 
\psi~:~(h,\hb)=(\hff,0) \qquad \psib~:~(h,\hb)=(0,\hff)
\EE
Thus, the mass term $\psi^\dg\psib-\psib^\dg\psi$ has scaling dimension
$\Delta=1$ and zero conformal spin; according to Eq.~(\ref{gap}), this
term gives rise to a gap proportional to $m$, as expected of course. On
the other hand, the perturbation $\psi^\dg\psi + \psib^\dg\psib$ , even if
it also has scaling dimension 1, is a sum of terms with nonzero conformal spin.
For such terms the analysis leading to Eq.~(\ref{gap}) cannot be applied.
Perturbations with conformal spin, such as this one, will typically
change the slow wavevectors, cause incommensurability, etc.

\begin{figure}
\figEPS{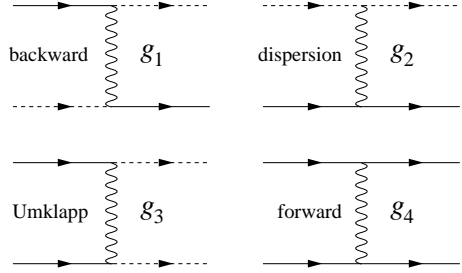}{6cm}
\caption{The four scattering processes for right-moving (continuous
lines) and left-moving (dashed lines) electrons in one dimension: (i)
$g_4$ is the amplitude for forward scattering; (ii) $g_2$ for dispersion
(scattering of left onto right electrons); (iii) $g_1$ for backward
scattering (each electron changes branch, but keeps its spin); (iv) $g_3$ for
Umklapp scattering (two left electrons become right electrons or vice-versa;
possible only at half-filling because of crystal momentum conservation).
Spin indices are suppressed, but incoming electrons must have opposite
spins for $g_1$ and $g_3$.
}
\label{g-ology}
\end{figure}

\subsection{Interactions}

The general two-body interaction between 1D electrons (of one band) is
\BE
H_{\rm int.} = \hf\sum_{\s,\s'}\sum_{k_1,k_2,k_3} V_{\s,\s'}(k_1,k_2,k_3) 
c_\s^\dg(k_1)c_{\s'}^\dg(k_2)c_{\s'}(k_3)c_\s(k_4)
\EE
where we have restored spin indices.
In the low-energy limit, scattering processes are naturally restricted to the
vicinity of the Fermi points, and fall into four kinematic types, illustrated
on Fig.~\ref{g-ology}. In terms of continuum fields, the corresponding
Hamiltonian densities are:
\BEA
\Ham_1 &=& vg_1\sum_\s \psi_\s^\dg \psib_\s \psib_{-\s}^\dg \psi_{-\s} \cr
\Ham_2^c &=& vg_{2,c} (J_\up+J_\dn)(\Jb_\up+\Jb_\dn) \cr
\Ham_2^s &=& vg_{2,s}(J_\up-J_\dn)(\Jb_\up-\Jb_\dn) \cr
\Ham_3 &=& \hff vg_3\sum_\s \psi_\s^\dg \psi_{-\s}^\dg \psib_\s
\psib_{-\s} +{\rm H.c.} \cr
\Ham_4^c &=& \hff vg_{4,c} \l[(J_\up+J_\dn)^2+(\Jb_\up+\Jb_\dn)^2\r]\cr
\Ham_4^s &=& \hff vg_{4,s}
\l[(J_\up-J_\dn)^2+(\Jb_\up-\Jb_\dn)^2\r]
\EEA
Remarks:
\BI
\item The incoming spins in $\Ham_1$ can be assumed to be antiparallel, since
the case of parallel spins reduces to a density-density coupling of the form
$\Ham_2^s$.
\item We have supposed that the couplings $g_{1-4}$ are momentum
independent, and this can be safely assumed in the low-energy limit. However,
they are cutoff dependent, i.e., they are subject to a RG flow.
\item The interactions $\Ham_1$ and $\Ham_2^s$ are not separately spin-rotation
invariant, but their combination is if $g_1=-2g_{2,s}$. This will be shown in
subsection~\ref{NabosTLSS}.
\item That $g_{2,\mu}$ and $g_{4,\mu}$ are density-density interactions is made
more explicit in this real-space representation. If these couplings alone
are nonzero (i.e., $g_1=g_3=0$), the resulting model is called the
{\it Tomonaga-Luttinger model} and can be solved exactly by bosonization. 
\EI
The bosonized form of these interactions will be derived in
Sect.~\ref{bosointSS}.

\section{Bosonization: A Heuristic View}

\subsection{Why is one-dimension special?}

\begin{figure}
\figEPS{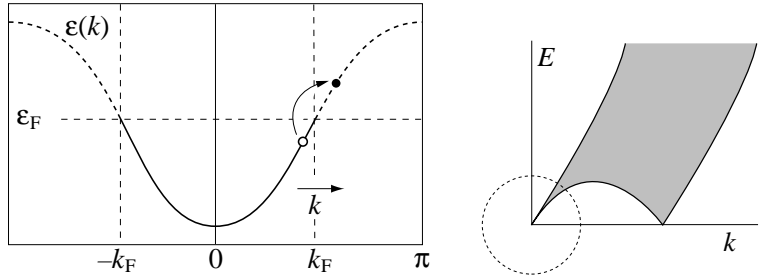}{10cm}
\caption{Particle-hole excitations in one dimension.
}
\label{PT-1DFIG}
\end{figure}

The basic idea behind bosonization is that particle-hole excitations are
bosonic in character, and that somewhat the greatest part (if not the
totality) of the electron gas spectrum might be exhausted by
these excitations. The question was raised by F.~Bloch back in 1934, but
Tomonaga realized in 1950 that it could only be true in one dimension. The
reason is simple: Consider Fig.~\ref{PT-1DFIG}. On the left the creation of
a particle-hole pair of momentum $k$ is illustrated.
On the right is the spectrum of
particle-hole excitations created out of the Fermi sea: the energy of the pair
is plotted vs the momentum of the pair (both quantities measured w.r.t. the
ground state). Because of the linear one-particle dispersion near the Fermi
level, the pairs have a narrow, quasiparticle-like dispersion near zero
momentum: they can propagate coherently. In other words, the particle and hole
have nearly the same group velocity at low energies and propagate together.
Any weak particle-hole attraction is then bound to have dramatic effects, i.e.,
to bind the pair into a coherently propagating entity: a new particle.

\begin{figure}
\figEPS{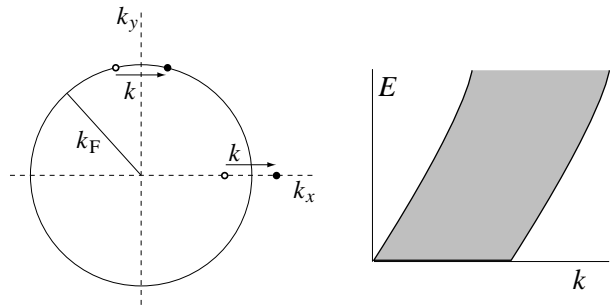}{8cm}
\caption{Particle-hole excitations in two dimensions.
}
\label{PT-2DFIG}
\end{figure}

On Fig.~\ref{PT-2DFIG}, the corresponding situation in two dimensions is
illustrated, with a circular Fermi surface. There it is clear that a
particle-hole pair with a given momentum $k$ can have a continuous spectrum of
energies, starting from zero. Thus, the particle-hole spectrum is a continuum
throughout and interactions have a harder time forming coherently propagating
particle-hole pairs.\footnote{This argument is partially weakened by the
presence of nesting.} In any case, defining the theory in terms of bosonic
excitations is much less obvious in dimensions greater than one.

In the low-energy limit, particle-hole excitations in on dimension are exactly
coherent and thus the density fluctuation fields $J_\s$ and $\Jb_\s$ create
propagating particles. The question is then whether any operator can be
expressed in terms of $J_\s$ and $\Jb_\s$, especially a single-particle creation
or annihilation operator. The answer is yes: the right-moving fermion field may
actually be expressed as
\BE
\psi_\s(x) = {1\over\sd} \exp\l[2\pi i\int^x \dr x'\; J_\s(x')\r] 
\EE
The rest of this section, and much of the following one, will be devoted to
proving Eq.~(\ref{bosoform2}), equivalent to the above.

\subsection{The simple boson}

Before going any further one must define what a free boson is. This may be done
fundamentally from a Lagrangian. For a massless boson field $\p$, this is
\BE\label{boseLag1}
L_0 = {1\over2}\int \dx \l[{1\over v}(\d_t\p)^2 - v(\d_x\p)^2 \r]
\EE
where $v$ is the speed of the Bose particles. The
corresponding Hamiltonian density is then
\BE\label{boseHamil1}
\Ham_0 = \hf v \l[\Pi^2 + (\d_x\p)^2 \r]
\qquad \Pi={1\over v}\d_t\p
\EE
where $\Pi$ is the field conjugate to $\p$, the two fields obeying the canonical
commutation rules
\BE
[\p(x),\Pi(x')] = i\delta(x-x')\qquad 
\BA{rl}[\p(x),\p(x')] &=0 \cr  [\Pi(x),\Pi(x')] &= 0\EA
\EE

The standard mode expansion for the fields $\p$ and $\Pi$ on the infinite line
is
\BEA\label{bosmod0}
\p(x)&=& \int \dk k \sqrt{v\over 2\om(k)}
\l[ b(k) \e^{ikx}+b^\dg(k) \e^{-ikx} \r] \NN\\
\Pi(x)&=& \int \dk k \sqrt{\om(k)\over 2v}
\l[-i b(k) \e^{ikx}+ib^\dg(k) \e^{-ikx} \r] 
\EEA
where $\om(k)\equiv v|k|$. The creation and annihilation operators obey the
commutation rules
\BE
[b(k),b^\dg(k')] = 2\pi\delta(k-k')
\EE
and the Hamiltonian may be expressed as
\BE
H = \int \dk k \om(k) b^\dg(k) b(k)
\EE
The ground state (or vacuum) $|0\R$ is annihilated by all the $b(k)$'s, and has
zero energy by convention. 

Like fermions, Bose particles in 1D either move to the left or to the right.
Indeed, the time dependence of the annihilation operators is obtained
through multiplying by the phase $\e^{-iv|k|t}$, if the boson is massless. A
separation into left- and right-moving parts is then possible. We write
\BE\label{LRbos1}
\p(x,t) = \phi(x-vt) + \phib(x+vt)
\EE
where
\BEA
\phi(z) = \int_{k>0} \dk k {1\over\sqrt{2k}}
\l[ b(k) \e^{-kz}+b^\dg(k) \e^{kz} \r] \NN\\
\phib(\zb) = \int_{k>0} \dk k {1\over\sqrt{2k}}
\l[ \bb(k) \e^{-k\zb}+\bb^\dg(k) \e^{k\zb} \r]
\EEA
The variables $z$ and $\zb$ are defined as in Eq.~(\ref{zzb}). Note that
wavevectors now take only positive values, since we have defined $\bb(k)\equiv
b(-k)$.

\subsubsection*{The dual field}

It is customary to introduce the so-called {\it dual boson} $\dual(x,t)$,
defined by the relation $\d_x\dual= -\Pi= -{1\over
v}\d_t\p$. In terms of right and left bosons $\phi$ and $\phib$, this becomes
(cf Eq.~(\ref{zzb2}))
\BEA
\d_x\dual &=& -{1\over v}\d_t \p \NN\\
&=& -i(\d_z+\d_\zb) (\phi+\phib) \NN\\
&=& -i(\d_z-\d_\zb) (\phi-\phib) \NN\\
&=& \d_x(\phi-\phib)
\EEA
therefore $\t=\phi-\phib$, modulo an additive constant which we set to zero.
One may then write
\BE\label{phiphibdual}
\phi = \hf(\p+\dual) \qquad \phib = \hf(\p-\dual)
\EE
if one desires a definition of $\phi$ and $\phib$ that stands independent from
the mode expansion. Since $\dual$ is expressed nonlocally in terms of $\Pi$
(i.e., through a spatial integral), the above expression makes it clear that
the right and left parts $\phi$ and $\phib$ are not true local fields relative
to $\p$. The basic definition of $\dual$ in terms of $\Pi$ implies the
equal-time commutation rule
\BE
[\p(x),\dual(x')] = -i\t(x-x')
\EE
(please note that $\t(x)$ is the step function [$\ne\dual$])
which further shows the nonlocal relation between $\p$ and $\dual$ (the
commutator is nonzero even at large distances).

\subsection{Bose representation of the fermion field}

After this introduction of the boson field, let us proceed
with a heuristic (and incomplete) derivation of the bosonization formula. The
basic idea is the following (we ignore spin for the moment): the electron
density $\ntot$ being bilinear in the electron fields, it has Bose
statistics. Let us suppose that it is the derivative of a boson field: 
$\ntot\propto\d_x\p$. Then
\BE
\p(x) = \la \int_x^\infty \dy \ntot(y) \qquad 
\ntot(x) = -{1\over\la}\d_x\p(x)
\EE
where $\la$ is a constant to be determined.
Creating a fermion at the position $x'$ increases $\p$ by $\la$ if $x<x'$.
This has the same effect as the operator
\BE
\exp \l[ -i\la\int_{-\infty}^{x'} \dy \Pi(y) \r]
\EE
which acts as a shift operator for $\p(x)$ if $x<x'$, because of the
commutation relation
\BE
\l[ \p(x),\int_{-\infty}^{x'} \dy \Pi(y) \r] =  \l\{
\BA{ll} i\qquad&(x<x')\cr 0 &(x>x')\EA \r.
\EE
Thus, a creation
operator $\psi^\dg(x')$ or $\psib^\dg(x')$ may be represented by the above
operator, times an operator that commutes with $\p$ ($\e^{i\a\p}$,
for instance). This extra factor must be chosen such that $\psi^\dg$ depends on
$x-vt$ only, and $\psib^\dg$ on $x+vt$ only. The left-right decomposition
(\ref{LRbos1}) and Eq.~(\ref{phiphibdual}) are useful here. Since
\BE
\Pi={1\over v}\d_t\p = -\d_x\dual
\EE
we have
\BE
-i\la\int_{-\infty}^{x'} \dy \Pi(y) = i\la\dual(x')
\EE
from which purely right- and left-moving fields may be
obtained by adding or subtracting $-\hf i\a\p(x')$. Thus, a sensible Ansatz for
a boson representation of electron creation operators is
\BE\label{bosoform1}
\psi^\dg(x) = A \e^{2i\la\phi(x)} \qquad 
\psib^\dg(x) = A \e^{-2i\la\phib(x)}
\EE
where the constants $A$ and $\la$ are to be determined, for instance by imposing
the anticommutation relations (\ref{anticomm1}). This will be done below.

The one-electron states $\psi^\dg|0\R$, being generated by exponentials of
boson creation operators, are {\it coherent states} of the boson field
$\phi$.\footnote{The interactions $g_1$ and $g_3$ give the
boson Hamiltonian the sine-Gordon form (cf. Sect.~\ref{bosointSS}). In that
case, one-electron states correspond to solitons (kinks) of the sine-Gordon
theory.} On the other hand, the elementary Bose excitations (the mesons, in
field-theory jargon) are collective density (or spin) fluctuations, i.e.,
electron-hole excitations.

\section{Details of the Bosonization Procedure}
\label{bosoS}

\subsection{Left and right boson modes}

The mode expansions (\ref{bosmod0}) may seem reasonable, but are actually
incomplete, for the following reasons:
\BI
\item The field being massless, the mode
expansion is ill-defined at $k=0$, because $\om(k=0)=0$. The mode at $k=0$,
or {\it zero-mode}, has to be treated separately and turns out to be important.
The existence of this mode apparently spoils the left-right separation of the
field $\p$.
\item In order for bosonization
to be rigorously defined, the field $\p$ must have an angular character. In
other words, the target space of the field must be a circle of radius $R$ (to
be kept general for the moment), so that $\p$ and $\p+2\pi R$ are identified.
We say that the boson is {\it compactified} on a circle.
\item A rigorous proof of bosonization procedure (at the spectrum level) is best
obtained on a system of finite size $L$ with periodic boundary conditions.
\EI

Taking the above remarks into account leads to the following improved
mode expansion for the boson field $\p(x,t)$:
\BE
\p(x,t) = q + {\pi_0 vt\over L} + {\tilde\pi_0 x\over L}+
\sum_{n>0} {1\over\sqrt{4\pi n}}
\l[ b_n \e^{-kz}+b_n^\dg \e^{kz} +\bb_n \e^{-k\zb}+\bb_n^\dg \e^{k\zb} \r]
\EE
Wavevectors are now quantized as $k=2\pi n/L$ ($n$ an integer). The creation
and annihilation operators have a different normalization from
Eq.~(\ref{bosmod0}), so as to obey the commutation rules
\BE
[b_n,b_m^\dg] = \delta_{mn} \qquad
[\bb_n,\bb_m^\dg] = \delta_{mn}
\EE
The zero-mode is treated
explicitly as a pair of canonical variables $q$ and $\pi_0$ obeying the
commutation rule
\BE
[q,\pi_0] = i
\EE
Defining $\p$ on a circle of radius $R$ has two consequences:
\BI
\item The operator $q$ is not single-valued, since it is an angular
variable ($q$ and $q+2\pi R$ are identified). Only exponentials $\e^{inq/R}$,
where $n$ is an integer, are well defined. Because $\pi_0$ is conjugate to $q$,
we have the relation
\BE
\e^{-inq/R}\pi_0\e^{inq/R} = \pi_0 + {n\over R} 
\EE
(the exponential operators shifts the eigenvalues of $\pi_0$).
Starting with the $\pi_0=0$ ground state, this means that the spectrum of
$\pi_0$ is restricted to $\Zn/R$.
\item Winding configurations are allowed; the constant $\tilde\pi_0$ is
precisely defined as $\tilde\pi_0 = 2\pi Rm$, where $m$ is the number of
windings of $\p$ as $x$ goes from
$0$ to $L$. It turns out to be useful to introduce an {\it operator}
$\tilde\pi_0$, defined by its eigenvalues $2\pi Rm$. This operator commutes
with all other operators met so far, but one may define an operator $\tilde q$
conjugate to $\tilde\pi_0$, i.e., such that $[\tilde q,\tilde\pi_0]=i$. 
\EI

The separation between left- and right-moving parts may then be done in spite of
the existence of a zero mode:
\BEA\label{bosmode1}
\phi(z) &=& Q + {P\over 2L}(vt-x) + \sum_{n>0} 
{1\over\sqrt{4\pi n}} (b_n \e^{-kz}+b_n^\dg \e^{kz}) \NN\\
\phib(z) &=& \bar Q + {\bar P\over 2L}(vt+x) + \sum_{n>0} 
{1\over\sqrt{4\pi n}} (\bb_n \e^{-k\zb}+\bb_n^\dg \e^{k\zb}) 
\EEA
where we have defined the right and left zero-modes
\BEA
Q &=& \hf(q-\tilde q) \qquad P = \pi_0-\tilde\pi_0 \qquad [Q,P] =i \NN\\
\bar Q &=& \hf(q+\tilde q) \qquad \bar P = \pi_0+\tilde\pi_0 
\qquad [\bar Q,\bar P] =i 
\EEA
Note that the ``artificial'' operator $\tilde q$ drops from the combination
$\p=\phi+\phib$. The spectrum of $P$ and $\bar P$ is
\BE\label{spectrum1}
P = \l( {n\over R}+2\pi Rm \r) \qquad
\bar P = \l( {n\over R}-2\pi Rm \r)
\EE
$m$ and $n$ being integers. Note that this spectrum is not a simple Cartesian
product, which means that the left-right decomposition of the boson is not
perfect, despite the expansion (\ref{bosmode1}).

\subsection{Proof of the bosonization formulas: Vertex operators}

Before investigating the constants $A$ and $\la$ of Eq.~(\ref{bosoform1}),
serious questions must be asked on the precise definition of an exponential
operator of the form $\e^{i\a\phi}$ (such operators are called {\it vertex
operators}, from their usage in string theory). Because
$\phi$ is a fluctuating field, any power or more complicated function of $\phi$
requires a careful definition if it is not to have a divergent average value.
The natural prescription is called {\it normal ordering} and consists in
expressing $\phi$ as a mode expansion and putting all annihilation operators to
the right of creation operators. Thus, the exponential operator has the
following precise definition:
\BE
\no{\e^{i\a\phi(x-vt)}} =
\e^{i\a Q} \exp\l[i\a\sum_{n>0}{1\over\sqrt{4\pi n}} b^\dg_n \e^{kz}\r]
\exp\l[i\a\sum_{n>0}{1\over\sqrt{4\pi n}} b_n \e^{-kz} \r]
\e^{-i\a P(x-vt)/2L} 
\EE
where $P$ is considered an annihilation operator since the boson ground state
$|0\R$ has zero momentum. The notation $\no{A}$ means that the operator $A$ is
normal ordered.

Because of normal ordering, vertex operators do not multiply like ordinary
exponentials. Instead, one has the relation
\BE\label{vertex1}
\e^{i\a\phi(z)}\e^{i\b\phi(z')} = \e^{i\a\phi(z)+i\b\phi(z')}
\e^{-\a\b\L\phi(z)\phi(z')\R}
\EE
where $\L\phi(z)\phi(z')\R$ is a simple ground state expectation value if the
left-hand side is an ordinary product, or a Green function if it is a
time-ordered product. This is one of the most important formulas of this
review, and it can be easily demonstrated using the Campbell-Baker-Hausdorff
(CBH) formula:
\BE\label{CBH}
\e^A \e^B = \e^{A+B}\e^{\hf[A,B]} \qquad([A,B]={\rm const.})
\EE
Consider for instance a single harmonic oscillator $a$, and two operators
\BE
A = \a a+\a' a^\dg \qquad B = \b a+\b' a^\dg
\EE
The CBH formula allows for a combination of the normal-ordered exponentials:
\BEA
\no{\e^A} \no{\e^B} &=& \e^{\a' a^\dg}\e^{\a a}\e^{\b' a^\dg}\e^{\b a} \NN\\
&=& \e^{\a' a^\dg}\e^{\b' a^\dg}\e^{\a a}\e^{\b a}\e^{\a\b'} \NN\\
&=& \no{\e^{A+B}} \e^{\L 0|AB|0\R}
\EEA
The last equality also applies if $(a,a^\dg)$ is replaced, in the same order, by
a pair $(p,q)$ of canonically conjugate operators ($[q,p]=i$). It also applies
to a combination of independent oscillators, such as the boson fields $\phi$ or
$\phib$. Finally, it applies to a time-ordered product as well as to an
ordinary product, as long as $\L 0|AB|0\R$ is also replaced by a time-ordered
product. Thus, the identity (\ref{vertex1}) is proven.

It remains to calculate the Green function $\L\phi(z)\phi(z')\R$.
This can be done in several ways. For instance, by using the mode expansion
(\ref{bosmode1}) and taking the limit $L\to\infty$:
\BEA
\L\phi(z)\phi(0)\R &=& -{i\over 2L}(vt-x) + {1\over 4\pi}
\sum_{n>0}{1\over n}\e^{-2\pi nz/L} \NN\\
&=& -{z\over 2L} - {1\over 4\pi}\ln{2\pi z\over L} \NN\\
&\to& - {1\over 4\pi}\ln z + {\rm const.} \qquad(L\to\infty)
\EEA
We will adopt the normalization
\BE\label{phicorr}
\L\phi(z)\phi(z')\R = -{1\over 4\pi}\ln(z-z')
\EE
and drop the constant term, which basically defines an overall length scale.
Likewise, one finds
\BE\label{boscorrR}
\L\phib(\zb)\phib(\zb')\R = -{1\over 4\pi}\ln(\zb-\zb')
\EE
Another way of computing the boson Green function is to notice that the
nonchiral Green function ${\cal G}(x,\tau)=\L \p(x,\tau)\p(0,0)\R$ must obey
the two-dimensional Poisson equation
\BE
\nabla^2 {\cal G} = \delta(x)\delta(v\tau)
\EE
in Euclidian space-time. The solution
to that equation is readily obtained in polar coordinates, up to an additive
constant (the correct normalization is obtained by applying Gauss' theorem
around the origin):
\BE
{\cal G}(x,v\tau) =  -{1\over 4\pi}\ln(x^2+v^2\tau^2)
=-{1\over 4\pi}\ln(z\zb)
\EE
The result is thus equivalent to what can be inferred from the mode expansion,
since
\BE
\L \p(x,\tau)\p(0,0)\R = \L\phi(z)\phi(0)\R+\L\phib(\zb)\phib(0)\R
\EE
Formula (\ref{vertex1}) can thus be rewritten as
\BE\label{vertex2}
\e^{i\a\phi(z)}\e^{i\b\phi(z')} = \e^{i\a\phi(z)+i\b\phi(z')}
(z-z')^{\a\b/4\pi}
\EE

We are now in a position to demonstrate the boson-fermion equivalence at the
level of the (anti)commutation relations. In fact, it is simpler to
determine the constants $A$ and $\la$ of Eq.~(\ref{bosoform1}) by comparing
Green functions rather than simply looking at equal time (anti)commutators
(this avoids the equal time singularity). The electron propagator is readily
calculated from the mode expansion (\ref{fermimode1}):\footnote{For a
calculation in the path-integral formalism, see Ref.~\cite{DMS97}.}
\BEA\label{greenf1}
\L \psi(z)\psi^\dg(z')\R &=& 
\int_{k>0}\dk k \int_{q>0}\dk q \L 0|\a(k)\beta^\dg(q)|0\R \e^{-kz+qz'} \NN\\
&=& \int_{k>0}\dk k \e^{-k(z-z')} \NN\\
&=& {1\over2\pi}{1\over z-z'}
\EEA
where we have supposed that $\tau>\tau'$, which garantees the convergence of
the integral and follows from the prescribed time-ordered product.
Likewise,
\BE
\L \psi^\dg(z)\psi(z')\R = {1\over2\pi}{1\over z-z'}
\EE
The correct anticommutator may be obtained from the equal time limit:
\BEA\label{anticomm}
\{\psi(x,0),\psi^\dg(x',0)\} &=& \lim_{\eps\to0}\l(
\L \psi(\eps-ix)\psi^\dg(-ix')\R + \L \psi^\dg(\eps-ix')\psi(-ix)\R \r) \NN\\
&=& {1\over 2\pi}\lim_{\eps\to0}\l(
{1\over \eps-i(x-x')}+{1\over \eps-i(x'-x)}\r) \NN\\
&=& {1\over 2\pi}\lim_{\eps\to0} {2\eps\over (x-x')^2+\eps^2} =
\delta(x-x')
\EEA
($\psi$ or $\psib$ appearing with a single argument are considered functions
of $z$ or $\zb$). One easily checks  that the space-time Green function
(\ref{greenf1}) has the expected expression in momentum-frequency space:
\BE
{1\over2\pi}{1\over z-z'} \longrightarrow {1\over \om-v|k|}
\EE

To demonstrate Eq.~(\ref{bosoform1}) and fix the values of $A$ and $\la$,
one needs to calculate
\BE
\L \psi(z)\psi^\dg(z')\R = A^2 \L \e^{-2i\la\phi(z)} \e^{2i\la\phi(z')}\R
\EE
From the relation (\ref{vertex2}), this is
\BEA\label{bosotemp1}
\L \psi(z)\psi^\dg(z')\R 
&=& A^2 \L \e^{-2i\la(\phi(z)-\phi(z'))}\R (z-z')^{-\la^2/\pi} \NN\\
&=& A^2 (z-z')^{-\la^2/\pi} 
\EEA
Note that the normal-ordered expectation value
\BE
\L \e^{-i\a\phi(z)-i\b\phi(z')}\R
\EE
is nonzero only if $\a+\b=0$. This can be inferred from the action of the
zero-mode operator $Q$:\footnote{It is also a consequence of the U(1)
symmetry $\p\to\p+a$, where $a$ is a constant.}
\BE
\L 0| \e^{i(\a+\b)Q} \e^{-i(\a+\b)P/2L}|0\R
\EE
vanishes if $\a+\b\ne0$, since the exponential of $Q$ acting on $|0\R$ yields
another eigenstate of $P$, with eigenvalue $\a+\b$, orthogonal to $|0\R$.
The condition $\a+\b=0$ is called the neutrality condition.\footnote{This
terminology comes from an analogy with the Coulomb gas system of
two-dimensional statistical mechanics.} By comparing Eq.~(\ref{bosotemp1})
with Eq.~(\ref{greenf1}), we conclude that the constant $\la$ must be $\sqrt\pi$
and that
$A$ must be $1/\sd$. The correct bosonization formula is then
\BE\label{bosoform2}
\BA{rlrl}
\psi(x) &= \Frac1\sd \e^{-i\sk\phi(x)}  \qquad 
&\psi^\dg(x) &= \Frac1\sd\e^{i\sk\phi(x)} \\[12pt]
\psib(x) &= \Frac1\sd \e^{i\sk\phib(x)}  
&\psib^\dg(x) &= \Frac1\sd \e^{-i\sk\phib(x)}\\
\EA\EE
Notice the change of sign in the exponent between $\psi$ and $\psib$.
Notice also that the choice of additive constant leading to Eq.~(\ref{boscorrR})
influences the value of $A$ only, the end result being the same irrespective of
that choice.

\noindent Remarks:
\BI
\item It is also instructive to check the anticommutation relation
\BE
\{\psi(x),\psi(x')\}=0
\EE
This is done simply by noting that the Green function
\BE
\L \psi(z)\psi(z')\R= {1\over2\pi}
\L\e^{-i\sk\phi(z)}\e^{-i\sk\phi(z')}\R
\EE
vanishes, by the neutrality condition.
The anticommutator, obtained in the equal-time limit, also vanishes.
\item The vertex operator $\e^{i\a\phi}$ is a scaling field. Its conformal
dimension $h_\a$ may be retrieved from the exponent of the Green function
\BE
\L \e^{i\a\phi(z)}(\e^{i\a\phi(0)})^\dg\R = {1\over z^{\a^2/4\pi}}
\EE
and from the relation~(\ref{2point}). The same analysis may be applied
to a left-moving vertex operator $\e^{i\bar\a\phib}$. Thus,
\BE\label{dimvertex}
\BA{rlrl}
h(\a) &= \Frac{\a^2}{8\pi} \qquad &\hb(\a) &=0 \\
h(\bar\a) &= 0 \qquad &\hb(\bar\a) &=\Frac{\bar\a^2}{8\pi}
\EA
\EE
Roughly speaking, the boson field has scaling dimension zero and its powers
series may produce operators with nonzero scaling dimension because of the
necessary normal ordering.
\EI

\subsection{Bosonization of the free-electron Hamiltonian}

The bosonization formulas (\ref{bosoform2}) are the basis of a 
fermion-to-boson translation of
various operators. However, this translation process is often
subtle, since normal ordering of these operators is required. Take for
instance the chiral fermion densities $J$ and $\Jb$. Normal ordering
can be constructed from the mode expansion. But a shorter and more elegant way
to achieve it is by {\it point splitting}, i.e., by defining
\BE
J(z) = \lim_{\eps\to0} \l[ \psi^\dg(z+\eps)\psi(z) 
- \L\psi^\dg(z+\eps)\psi(z)\R \r]
\EE
where $\eps$ has positive time (so as to appear as shown after time ordering).
This limit must be taken after applying the bosonization formulas and the
relation (\ref{vertex1}):
\BEA\label{OPE1}
J &=& {1\over 2\pi}\lim_{\eps\to0}\l[
\e^{i\sk\phi(z+\eps)}\e^{-i\sk\phi(z)}-{1\over \eps} \r] \NN\\
&=& {1\over 2\pi}\lim_{\eps\to0}\l[
\e^{i\sk[\phi(z+\eps)-\phi(z)]}{1\over\eps} - {1\over \eps} \r] \NN\\
&=& {1\over 2\pi}\lim_{\eps\to0}\l[
\e^{i\eps\sk\d_z\phi(z)}{1\over\eps} - {1\over \eps} \r] \NN\\
&=& {1\over 2\pi} i\sk \d_z\phi = {i\over\sqrt\pi} \d_z\p
\EEA
where we have Taylor-expanded the exponential on the last line.
We proceed likewise for the left-moving density, except that the sign is
opposite. To summarize:
\BE\label{U1currents}
J = {i\over\sqrt\pi} \d_z\p \qquad
\Jb = -{i\over\sqrt\pi} \d_\zb\p
\EE
(note that $\d_z\phi=\d_z\p$ and $\d_\zb\phib=\d_\zb\p$).

The density fluctuations $J$ and $\Jb$ are also components of a conserved
current from the boson point of view. Indeed, the compactified boson has a
$U(1)$ symmetry: the Lagrangian is invariant under the shift $\p\to\p+a$. By
Noether's theorem, this symmetry implies the existence of a conserved quantity,
with a density proportionnal to $\d_t\p$ and current proportionnal to
$v^2\d_x\p$. It is customary to define right and left `currents' exactly as
in Eq.~(\ref{U1currents}), so that the continuity equation reduces to
$\d_z\Jb+\d_\zb J=0$. In fact, the critical (i.e., massless) and
one-dimensional nature of the theory enhance this $U(1)$ symmetry to a {\it
chiral} $U(1)$ symmetry, by which the left and right currents are separately
conserved:
\BE
\d_\zb J = 0 \qquad \d_z\Jb=0
\EE

The same short-distance expansion as in Eq.~(\ref{OPE1}) may be used to
demonstrate explicitly the equivalence of the electron Hamiltonian
(\ref{contHam2}) with the boson Hamiltonian (\ref{boseHamil1}), except that a
higher-order expansion is necessary. More explicitly, an expression like
$\psi^\dg\d_x\psi$ must be evaluated as
\BE\label{limit1}
\psi^\dg\d_x\psi = -i\lim_{\eps\to 0}\l\{
\psi^\dg(z+\eps)\d_z\psi(z) -\L\psi^\dg(z+\eps)\d_z\psi(z) \R\r\}
\EE
The limit is evaluated by using the following short-distance products (to order
$\eps^2$), derived from the bosonization formulas and the relation
(\ref{vertex1}):
\BEA
\psi(z')\psi^\dg(z) &=& {1\over 2\pi\eps} +
i{1\over\sqrt\pi}\d_z\p + i{\eps\over\sk}\d^2_z\p  -
\eps(\d_z\p)^2 \NN\\
\psi^\dg(z')\psi(z) &=&
{1\over 2\pi\eps} - i{1\over\sqrt\pi}\d_z\p
- i{\eps\over\sk}\d^2_z\p - \eps(\d_z\p)^2\NN\\
\psib(\zb')\psib^\dg(\zb) &=&
{1\over 2\pi\epsb} - i{1\over\sqrt\pi}\d_\zb\p-
i{\epsb\over\sk}\d^2_\zb\p - \epsb(\d_\zb\p)^2 \NN\\
\psib^\dg(\zb')\psib(\zb) &=&
{1\over 2\pi\epsb} + i{1\over\sqrt\pi}\d_\zb\p
+ i{\epsb\over\sk}\d^2_\zb\p -
\epsb(\d_\zb\p)^2 \quad
\EEA
where $\eps=z'-z$ and $\epsb=\zb'-\zb$, and the fields on the r.h.s. are
considered functions of $(z,\zb)$ only, the $(z',\zb')$ dependence residing
in the powers of $\eps$ and $\epsb$ only. The limit (\ref{limit1}) may be
obtained by differentiating with respect to $z$ or $\zb$ the above expansions,
and proceeding to a straightforward substitution. The Hamiltonian
(\ref{contHam2}) then becomes
\BEA\label{HU1}
H_{\rm F} &=& -v\int\dx \l[(\d_z\p)^2+(\d_\zb\p)^2\r] \NN\\
&=& \pi v\int\dx \l( J^2 + \Jb^2 \r)
\EEA
which is precisely the boson Hamiltonian (\ref{boseHamil1}), after using the
relations (\ref{zzb2}). Thus the equivalence between a free fermion and a free
boson is demonstrated at the level of the Hamiltonian. Note that we have not
demonstrated this equivalence at the Lagrangian level; this cannot be done with
the techniques used here.

\subsection{Spectral equivalence of boson and fermion}

That the free boson (at a certain radius $R$) is equivalent to a free fermion
requires that the spectra of the two theories be identical. This can be
tested by comparing the grand partition functions of the two models. Let us try
to do it for left and right excitations separately. Consider the right-moving
fermion $\psi(x)$, and the corresponding Hamiltonian and fermion number:
\BE
H_{\rm F} = \sum_{k>0} vk(\a_k^\dg\a_k + \b_k^\dg\b_k) \qquad
N_{\rm F} = \sum_{k>0} (\a_k^\dg\a_k - \b_k^\dg\b_k)
\EE
where the operators $\a_k$ and $\b_k$ have different normalizations from
those of Eq.~(\ref{contHam1}) and obey the anticommutation rules
\BE
\{\a_k,\a_q^\dg\} = \delta_{kq} \qquad \{\a_k,\a_q \} = 0
\EE
(and likewise for $\b_k$).
The spectrum of the Hamiltonian is encoded in the grand partition function
\BE
{\cal Z}_{\rm F} = \sum_{\rm states} \e^{-\b(H-\mu N)}
\EE
We have defined the theory on a cylinder, so as to have discrete wavenumbers.
We shall use antiperiodic boundary
conditions, known in string theory as {\it Neveu-Schwarz boundary conditions}:
\BE
\psi(x+L) = -\psi(x)
\EE
The wavenumbers are then half-integer moded: $k=2\pi(n+\hf)/L$, where $n$
is a positive integer (by contrast, periodic boundary conditions, also known as
{\it Ramond boundary conditions}, lead to integer-moded wavenumbers and to the
presence of a fermion zero-mode). Each mode being independent, the fermion grand
partition function factorizes as
\BE
{\cal Z}_{\rm F} = \prod_{n=1}^\infty (1+q^{n+1/2} t)(1+q^{n+1/2}t^{-1})
\qquad q\equiv \e^{-2\pi v\beta/L}~,~t\equiv \e^{\beta\mu}
\EE

Let us now turn to the boson. We first have to write down an expression
for the Hamiltonian in terms of the modes appearing in (\ref{bosmode1}).
A straightforward calculation yields
\BE
H_{\rm B} = {P^2\over 4L} + {2\pi\over L}\sum_{n>0} n b_n^\dg b_n
\EE
for the right-moving part. Let us then notice that the fermion number
$N_{\rm F}$ is proportional to the zero-mode $P$. More precisely,
\BE
N_{\rm F} = {P\over\sk}
\EE
This can be shown from the mode expansion (\ref{bosmode1}) and by the
equivalence $J=-\d_x\phi/\sqrt\pi$, which implies
\BE
N_{\rm F} = \int\dx J = -{1\over\sqrt\pi}(\phi(L)-\phi(0)) = {P\over\sk}
\EE
Each mode $b_n$ being independent, the boson grand partition function is then
\BEA
{\cal Z}_{\rm B} &=& \l\{\prod_{n=1}^\infty\l( 1+
q^n+q^{2n}+q^{3n}+\cdots\r)\r\}
\sum_P q^{P^2/8\pi}t^{P/\sk}\NN\\
&=& \l\{\prod_{n=1}^\infty{1\over 1-q^n}\r\}
\sum_P q^{P^2/8\pi}t^{P/\sk}
\EEA
So far we have not determined the boson radius $R$, nor the spectrum of $P$.
We now invoke Jacobi's triple product formula:
\BE\label{jacobi}
\prod_{n=1}^\infty (1-q^n)(1+q^{n-1/2}t)(1+q^{n-1/2}t^{-1})=
\sum_{n\in\Zn} q^{n^2/2}t^n
\EE
Applied to the fermion partition function, this remarkable identity shows that
\BE
{\cal Z}_{\rm F} = \l\{\prod_{n=1}^\infty{1\over 1-q^n}\r\}
\sum_n q^{n^2/2}t^n
\EE
In order to have ${\cal Z}_{\rm F}={\cal Z}_{\rm B}$, we therefore need
$P$ to take the values $P = n\sk$. This is not quite the spectrum
(\ref{spectrum1}). The closest it comes to is obtained at radius $R=1/\sk$,
where (\ref{spectrum1}) becomes
\BE\label{spectrum2}
P = \sk(n+\hf m) \qquad \bar P = \sk(n-\hf m)
\EE
In fact, because of the imperfect left-right separation of the boson spectrum
(\ref{spectrum1}), the equivalence with a perfectly left-right separated
fermion theory is impossible. However, a state-by-state correspondence between
the two theories exists if some constraints are imposed on the fermion
spectrum. Explicitly:
\BI
\item One needs to include both periodic and antiperiodic
boundary conditions in the fermion theory: the Hilbert space is then the
direct sum of two sectors.
\item The periodic sector must contain a single
zero-mode $\psi_0$ (the same for left- and right-moving fermions), such that
$\{\psi_0,\psi_0^\dg\}=1$. The operator
$\psi_0^\dg$ does not change the energy of a state, but increases the fermion
number by one. This mixes (in a weak way) the left and right-moving spectra.
\item the number of fermions in the antiperiodic sector must be even (odd-number
states are thrown out).
\EI
These constraints will not be demonstrated here; a detailed proof can
be found in Refs \cite{DMS97} and \cite{Ginsparg89}.

That the simple-minded correspondence between left-moving fermions and
left-moving bosons is impossible should not bother us too much. The
complications mentioned above basically involve boson or fermion zero-modes and
boundary conditions. The nonzero frequency modes are not affected and the
fermion-boson correspondence works well for those modes, as reflected in the
Jacobi triple product formula (\ref{jacobi}). First and foremost, we should
retain from this exercise that the boson radius must be $R=1/\sk$ in order for
the fermion-boson equivalence to hold.

\subsection{Case of many fermion species: Klein factors}

Suppose now that we have more than one fermion species, labelled by a greek
index: $\psi_\mu$. Bosonization would then require an equal number of boson
species $\phi_\mu$ and the formulas (\ref{bosoform2}) would still be
applicable, except for the fact that they do not provide for the
anticommutation of different fermion species. In fact, this problem arises for
a single species as well, since the left and right fermions $\psi$ and $\psib$
must anticommute, whereas the fields $\phi$ and $\phib$ commute.\footnote{In our
periodic quantization scheme. On the infinite line, with vanishing boundary
conditions at $x=\pm\infty$, this is not true and $[\phi(x),\phib(x')]=\frac14
i$. This garantees the anticommutation $\{\psi(x),\psib(x')\}=0$.}
A possible solution lies in the introduction of additional anticommuting
factors, the so-called {\it Klein factors}, in the bosonization formulas:
\BE\label{bosoform3}
\BA{rlrl}
\psi_\mu(x) &= \Frac1\sd\eta_\mu \e^{-i\sk\phi_\mu(x)}  \qquad 
&\psi_\mu^\dg(x) &=\Frac1\sd \eta_\mu\e^{i\sk\phi_\mu(x)} \\[12pt]
\psib_\mu(x) &= \Frac1\sd\etab_\mu \e^{i\sk\phib_\mu(x)}  \qquad 
&\psib_\mu^\dg(x) &=\Frac1\sd \etab_\mu\e^{-i\sk\phib_\mu(x)} \\
\EA\EE
The Klein factors $\eta_\mu$ and $\etab_\mu$ are Hermitian
and obey the Clifford algebra:
\BE\label{cliff}
\{ \eta_\mu,\eta_\nu\} = \{ \etab_\mu,\etab_\nu\} = 2\delta_{\mu\nu}
\qquad \{ \eta_\mu,\etab_\nu\} = 0
\EE
Klein factors act on a Hilbert space distinct from
the boson Hilbert space generated by the modes of Eq.~(\ref{bosmode1}).
This Hilbert space expansion must be compensated by some sort of
``gauge fixing'': the Hamiltonian must be diagonal in this Klein factor
space, as well as required physical observables. Then, one Klein eigenstate
is chosen and the rest of the Klein Hilbert space decouples.

An alternative to Klein factors, which requires no extra Hilbert space, is to
include in the bosonization formula for $\psi_\mu$ the factor
\BE
\exp\l[ i\pi \sum_{\nu<\mu}N_{{\rm F},\nu} \r]
=\exp\l[i{\sqrt\pi\over2}\sum_{\nu<\mu}P_\nu \r]
\EE
($P_\nu$ is the conjugate to the zero mode $Q_\mu$, cf. Eq.~(\ref{bosmode1})).
Because of the commutation rule $[Q_\mu,P_\nu] = i\delta_{\mu\nu}$, the
CBH formula (\ref{CBH}) makes the different fermions species anticommute.
This must be properly extended to left-moving fermions, for instance by
enumerating them consecutively after the right fermions (unless one uses
the infinite-line quantization scheme).

\subsection{Bosonization of interactions}
\label{bosointSS}

Let us finally turn our attention to the basic interactions (\ref{g-ology}).
This requires introducing spin: we need two bosons $\p_\up$ and $\p_\dn$.
The boson fields may be naturally combined into spin and charge components:
\BE
\p_c = {1\over\sqrt2}(\p_\up+\p_\dn) \qquad
\p_s = {1\over\sqrt2}(\p_\up-\p_\dn)
\EE
and likewise for the chiral components $\phi_{c,s}$ and $\phib_{c,s}$.

\subsubsection*{Backward scattering}
The backward scattering term ($g_1$) does not need normal ordering, since
all four Fermi fields anticommute. A straightforward application of
the bosonization formulas is sufficient:
\BEA
\Ham_1 &=& {vg_1\over(2\pi)^2}\sum_\s \eta_\s\etab_\s\etab_{-\s}\eta_{-\s}
\exp\l[i\sk(\phi_\s+\phib_\s-\phib_{-\s}-\phi_{-\s})\r] \NN\\
&=& {vg_1\over(2\pi)^2}\l[
\eta_\up\etab_\up\etab_\dn\eta_\dn \e^{i\sqrt{8\pi}\p_s} 
+ \eta_\dn\etab_\dn\etab_\up\eta_\up \e^{-i\sqrt{8\pi}\p_s} \r]
\EEA
Since all the Klein factors present anticommute,
$\eta_\up\etab_\up\etab_\dn\eta_\dn=\eta_\dn\etab_\dn\etab_\up\eta_\up$
and it remains
\BE
\Ham_1 = {vg_1\over 2\pi^2}
\eta_\up\etab_\up\etab_\dn\eta_\dn \cos(\sqrt{8\pi}\p_s)
\EE
The product of Klein factors poses a slight conceptual problem.
The Klein Hilbert space (on which the $\eta_\s$ and $\etab_\s$ act)
is the representation space of a Clifford algebra, and has a minimal
dimension which depends on the number of fermion species. With four
species (two spins, times left and right), the minimal dimension is
four, exactly like Dirac matrices in four-dimensional space-time.
For instance, the following representation is possible, in terms of
tensor products of Pauli matrices:
\BE
\BA{rlrl}
\eta_\up = \s_1\otimes\s_1 \qquad &\eta_\dn &= \s_3\otimes\s_1 \\
\etab_\up = \s_2\otimes\s_1 \qquad &\etab_\dn &= 1\otimes\s_2 
\EA\EE
One readily checks that this is a faithful representation of the
Clifford algebra (\ref{cliff}). Then, the above Klein product
is
\BE\label{Klein1}
\eta_\up\etab_\up\etab_\dn\eta_\dn = 1\otimes\s_3
\EE
This matrix is diagonal with eigenvalues $\pm 1$. If no other term in the
Hamiltonian contains a nondiagonal Klein product, one may safely pick a Klein
eigenstate and stick with it, forgetting the rest of the Klein Hilbert space.
This is some kind of gauge choice.\footnote{One has to be careful when
calculating various correlation functions, since the operators involved may
also contain Klein factors, and simultaneous diagonality of all relevant
operators with interactions is important if one wants to forget about the
Klein Hilbert space.} In the case at hand, we may chose the eigenvalue $+1$ and
write
\BE\label{bosint1}
\Ham_1 = {vg_1\over 2\pi^2} \cos(\sqrt{8\pi}\p_s)
\EE

\subsubsection*{Forward scattering}
The next interaction term ($g_2$) is quite simple to bosonize, since it
involves only density operators:
\BE
\Ham_2^c = {vg_{2,c}\over\pi}\sum_{\s,\s'} \d_z\p_\s \d_\zb\p_{\s'}
= {2vg_{2,c}\over\pi} \d_z\p_c \d_\zb\p_c
\EE
and likewise for the spin interaction:
\BE
\Ham_2^s = {2vg_{2,s}\over\pi} \d_z\p_s \d_\zb\p_s
\EE
Likewise, the forward terms ($g_4$) are
\BEA
\Ham_4^c &=& -{vg_{4,c}\over2\pi}
\l[(\d_z\p_\up+\d_z\p_\dn)^2+(\d_\zb\p_\up+\d_\zb\p_\dn)^2\r] \NN\\
&=& -{vg_{4,c}\over\pi} \l[(\d_z\p_c)^2+(\d_\zb\p_c)^2\r]
\EEA
and
\BE
\Ham_4^s = -{vg_{4,s}\over\pi} \l[(\d_z\p_s)^2+(\d_\zb\p_s)^2\r]
\EE

\subsubsection*{Umklapp scattering}
Finally, the Umklapp term ($g_3$) is treated in the same way as $g_1$, this time
with the charge boson involved:
\BEA
\Ham_3 &=& {vg_3\over2(2\pi)^2}\l[
(\eta_\up \eta_\dn \etab_\up \etab_\dn +
\eta_\dn \eta_\up \etab_\dn \etab_\up)
\e^{i\sk(\phi_\up+\phi_\dn+\phib_\up+\phib_\dn)} + {\rm H.c.} \r] \NN\\
&=& {vg_3\over(2\pi)^2}\l[ \eta_\up \eta_\dn \etab_\up \etab_\dn
\e^{i\sqrt{8\pi}\p_c} + {\rm H.c.} \r] \NN\\
&=& {vg_3\over2\pi^2} \eta_\up \eta_\dn \etab_\up \etab_\dn
\cos(\sqrt{8\pi}\p_c)
\EEA
After a mere transposition, the Klein prefactor is the same as
(\ref{Klein1}). Therefore,
\BE\label{bosint3}
\Ham_3 = {vg_3\over2\pi^2} \cos(\sqrt{8\pi}\p_c)
\EE

\section{Exact solution of the Tomonaga-Luttinger model}

\subsection{Field and velocity renormalization}

The power of bosonization lies mainly in its capacity of representing
fermion interactions, namely $\Ham_2$ and $\Ham_4$, as parts of a noninteracting
boson Hamiltonian. The continuum electron model defined by the Hamiltonian
density
\BE
\Ham_{\rm T.L.} = \Ham_{\rm F} + \Ham_{2,c} + \Ham_{2,s} + \Ham_{4,c}
+ \Ham_{4,s} 
\EE
is called the {\it Tomonaga-Luttinger model} (TL model). 
The free electron part $\Ham_{\rm F}$ may be expressed as the sum of free boson
theories for the spin and charge bosons, with the same velocity $v$. As seen
in Sect.~\ref{bosointSS}, each interaction term is expressed either in terms of
the spin boson or of the charge boson. Thus, the total Hamiltonian is the sum
of charge and a spin Hamiltonians, i.e., the two sectors (charge and spin) are
completely decoupled:\footnote{This is true at the field level, but not
completely true at the spectrum level, where things are more subtle, as shown
above. Obviously the number of spin and charge excitations are not independent,
since the total number of electrons is conserved.}
\BE
\Ham_{\rm T.L.} = \Ham_c + \Ham_s 
\EE

When expressed in terms of the spin and
charge bosons -- more precisely, in terms of the conjugate field
$\Pi_\mu$ and of $\d_x\p_\mu$ -- the Hamiltonian of each sector becomes
\BEA\label{H2H4}
\Ham_\mu &=& \Ham_0^\mu + \Ham_2^\mu + \Ham_4^\mu \text{where}\NN\\
\Ham_{0,\mu} &=& \hf v \l[\Pi_\mu^2 + (\d_x\p_\mu)^2 \r] \NN\\
\Ham_{2,\mu} &=& -{vg_{2,\mu}\over2\pi}
\l[ \Pi_\mu^2 - (\d_x\p_\mu)^2 \r] \NN\\
\Ham_{4,\mu} &=&
{vg_{4,\mu}\over2\pi}\l[\Pi_\mu^2+(\d_x\p_\mu)^2\r] 
\EEA
where $\mu=c,s$. Combining these expressions, one finds
\BE\label{hamilK}
\Ham_\mu = \hff v_\mu \l[K_\mu\Pi_\mu^2 + {1\over K_\mu}(\d_x\p_\mu)^2 \r]
\EE
where
\BE\label{Ketv}
K_\mu = \sqrt{\pi-g_{2,\mu}+g_{4,\mu}\over \pi+g_{2,\mu}+g_{4,\mu}}
\qquad
v_\mu = v\sqrt{\l(1+{g_{4,\mu}\over\pi}\r)^2-\l({g_{2,\mu}\over\pi}\r)^2}
\EE
The constant $K_\mu$ is simply a renormalization of the field $\p_\mu$. This
is most obvious in the associated Lagrangian:
\BE
L_\mu = {1\over2K_\mu}\int \dx \l[{1\over v_\mu}(\d_t\p_\mu)^2 -
v_\mu(\d_x\p_\mu)^2 \r]
\EE
The Hamiltonian (\ref{hamilK}) may be brought back in the canonical form
(\ref{boseHamil1}) simply by introducing rescaled operators
\BE
\p_\mu' = {1\over\sqrt{K_\mu}}\p_\mu \qquad
\Pi_\mu' = \sqrt{K_\mu}\Pi_\mu 
\EE
which still obey the canonical commutation relations. Clearly, the field
$\p_\mu'$ has radius 
\BE
R_\mu ={R\over\sqrt{K_\mu}}={1\over\sqrt{4\pi K_\mu}}
\EE

\noindent Remarks:
\BI
\item
Let us point out that we have supposed the couplings $g_{2,\mu}$ and $g_{4,\mu}$ to be
momentum independent. However, this is not necessary: the above solution of the
Tomonaga-Luttinger model may be repeated with momentum-dependent $g_{2,\mu}(q)$ and
$g_{4,\mu}(q)$, for the simple reason that the boson Hamiltonian is quadratic and
that different momenta are therefore decoupled. In particular, the above
formula for the velocity renormalization may be used to write down the exact
dispersion relation of charge excitations:
\BE
\om(k) = v_c|k| =
v|k|\sqrt{\l(1+{g_{4,\mu}(k)\over\pi}\r)^2-\l({g_{2,\mu}(k)\over\pi}\r)^2}
\EE
For simplicity, we shall ignore from now on such a momentum dependence
of the couplings, mainly because it is generally irrelevant (in the
RG sense).

\item
It is important that the perturbations $\Ham_2$ and $\Ham_4$ of Eq.~(\ref{H2H4})
be considered as perturbations of the Hamiltonian, not of the Lagrangian. If the
conjugate momentum $\Pi$ were replaced by $\d_t\p/v$ and the resulting
expressions considered as perturbations of the Lagrangians, the field and
velocity renormalization would be quite different. The reason is that, as they
are expressed here, these perturbations depend on $\d_t\p_\mu$ and would
change the definition of the conjugate momentum, were they to be applied to 
the Langrangian.
\EI

\subsection{Left-right mixing}

The perturbations $\Ham_{2,\mu}$ of (\ref{H2H4}) mix right and left bosons.
One then expects that the new eigenstates will be linear combinations of left
and right excitations of the original boson. Indeed, rescaling the field $\p$
($\p'=\p/\sqrt K$) implies the opposite rescaling of its conjugate momentum
$\Pi$  -- and therefore of the dual field $\dual$ -- in order to preserve the
canonical commutation rules:
$\dual'=\dual\sqrt K$ (we drop the spin/charge index for the time being). Thus,
the left and right bosons $\phi$ and $\phib$ are not simply rescaled by
$K$, they are mixed:
\BEA
\phi &=& \hf(\p+\dual)\to \phi'=\hf\l({1\over\sqrt K}\p+\sqrt K\dual\r) \NN\\
\phib &=& \hf(\p-\dual)\to \phib'=\hf\l({1\over\sqrt K}\p-\sqrt K\dual\r)
\EEA
Expressing the old right and left bosons ($\phi$ and $\phib$) in terms of the
new ones ($\phi'$ and $\phib'$) and defining $\xi$ as $K=\e^{2\xi}$, one finds
\BEA
\phi  &=& \cosh\xi\;\phi' + \sinh\xi\;\phib' \NN\\
\phib &=& \sinh\xi\;\phi' + \cosh\xi\;\phib' 
\EEA
In terms of the mode expansion (\ref{bosmode1}), this left-right mixing
is a Bogoliubov transformation of the creation and annihilation
operators, if one collects like powers of $\e^{ikx}$ at equal times:
\BEA
b_n  &=& \cosh\xi\; b'_n + \sinh\xi\; \bar b'_n{}^\dg \NN\\
\bar b_n &=& \sinh\xi\; b'_n{}^\dg + \cosh\xi\;\bar b'_n
\EEA

Thus, the fermion field $\psi_\up$, a pure right-moving field in the
unperturbed system, becomes a mixture of right- and left-moving fields
in the Tomonaga-Luttinger model, if $K_\mu\ne1$ (i.e., if $g_{2,\mu}\ne0$).
Consider for instance the case $K_s=1$, $K_c\ne 1$:
\BEA\label{psibogo}
\psi_\up(x,\tau) &=& {1\over \sd}\eta_\up \e^{-i\sk\phi_\up} \NN\\
&=& {1\over \sd}\eta_\up
\e^{-i\sd\phi_c}\e^{-i\sd\phi_s} \NN\\
&=& {1\over \sd}\eta_\up
\e^{-i\sd\cosh\xi\;\phi'_c}\e^{-i\sd\sinh\xi\;\phib'_c}
\e^{-i\sd\phi_s}
\EEA
This is expected, since the interaction $\Ham_2$ surrounds the free electron
with a cloud of right {\it and} left electrons.

\subsection{Correlation functions}
\label{corrfuncSS}

\subsubsection*{Green function}
\label{GreenfuncSSS}

\begin{figure}
\figEPS{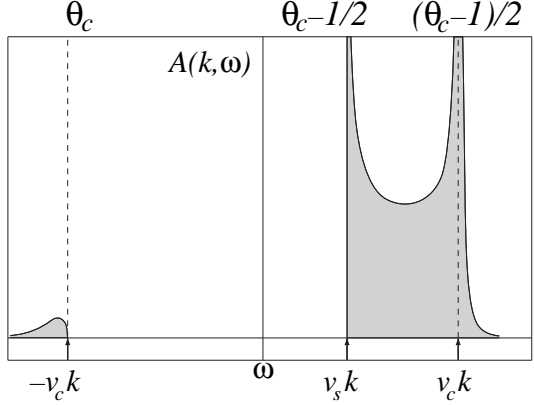}{7cm}
\caption{Typical spectral function $A(k,\om)$ of the Tomonaga-Luttinger
model, with $K_s=1$. The exponents appearing on top refer to the
three singularities: $(\om+v_c q)^{\t_c}$,
$(\om-v_s q)^{\t_c-1/2}$ and $(\om-v_c q)^{(\t_c-1)/2}.$
}\label{SpecLutt}
\end{figure}

Once the renormalizations (\ref{Ketv}) are obtained, the calculation of
physical quantities (correlation functions) is a straighforward
task in principle, at least in real space. Let us start with the
one-particle Green function
\BE
G_\up(x,\tau) = \L c_\up(x,\tau)c_\up^\dg(0,0)\R
\EE
Near zero wavevector, in terms of continuum fields, this
becomes\footnote{There is also a singular component near $k=2k_F$ is $K<1$.}
\BE
G_\up(x,\tau) = \L\psi_\up(x,\tau)\psi_\up^\dg(0,0)\R 
+ \L\psib_\up(x,\tau)\psib_\up^\dg(0,0)\R
\EE
Let us concentrate on the right-moving part, in the case $K_s=1$ and $K_c\ne1$
for simplicity. Using Eq.~(\ref{psibogo}), it becomes
\BEA
\L\psi_\up(x,\tau)\psi^\dg_\up(0,0)\R &=& {1\over 2\pi}
\L\e^{-i\sd\cosh\xi\;\phi'_c(z_c)}
\e^{i\sd\cosh\xi\;\phi'_c(0)}\R \NN\\
&&\times \L\e^{i\sd\sinh\xi\;\phib'_c(\zb_c)}
\e^{-i\sd\sinh\xi\;\phib'_c(0)}\R\NN\\
&&\times \L\e^{-i\sd\phi'_s(z_s)}\e^{i\sd\phi'_s(0)}\R \NN\\
&=& {1\over2\pi}{1\over (v_c\tau-ix)^{\hf\cosh^2\xi}}
{1\over (v_c\tau+ix)^{\hf\sinh^2\xi}}
{1\over (v_s\tau-ix)^{1/2}} \NN\\
&=& {1\over2\pi}{1\over (v_c\tau-ix)^{1/2}}
{1\over |v_c\tau-ix|^{\t_c}}
{1\over (v_s\tau-ix)^{1/2}}
\EEA
where the velocity difference between spin and charge demands a distinction
between spin and charge complex coordinates ($z_s$ and $z_c$). We have
introduced the exponent\footnote{The notation $\g_c=\hf\t_c$ is also used,
as is $\a_c=\t_c$.}
\BE
\t_c = \frac14 \l(K_c+{1\over K_c}-2\r)
\EE
Once the Fourier transform is taken in time and space, this Green function
leads to extended spectral weight,\footnote{For a detailed calculation of the
spectral weight from the above space-time Green function, see
Ref.~\cite{Gogolin98}} as illustrated on Fig.~\ref{SpecLutt}. This spectral
weight has power-law singularities near $\om=\pm v_cq$ and $\om=v_sq$, with
exponents related to $\t_c$. From this spectral function, it can be shown that
the momentum distribution function
$n(k)$ has an algebraic singularity at the Fermi level:
\BE
n(k) = n(k_F) - {\rm const.}\times{\rm sgn}(k-k_F)|k-k_F|^{\t_c} 
\EE
and so does the one-particle density of states:
$N(\om) \sim |\om|^{\t_c}$.

\subsubsection*{Compressibility}

Next, let us consider the compressibility near $q=0$:
\BE
\chi(x,\tau) = \L \ntot(x,\tau)\ntot(0,0) \R
\EE
Since $n_{\up,\rm tot.}=\d_x\p_\up/\sqrt\pi$,  we have
\BE
\ntot = \sqrt{2\over\pi} \d_x\p_c \text{whereas}
S_z = \sqrt{1\over2\pi} \d_x\p_s
\EE
is the $z$ component of the uniform spin density.
The compressibility is then
\BEA
\chi(x,\tau) &=& {2\over\pi}\L \d_x\p_c(x,\tau) \d_x\p_c(0,0)\R \NN\\
&=& {2K_c\over\pi}\L \d_x\p'_c(x,\tau) \d_x\p'_c(0,0)\R \NN\\
&=& -{2K_c\over\pi}\l[ \L \d_z\p'_c(x,\tau) \d_z\p'_c(0,0)\R 
+ \L \d_\zb\p'_c(x,\tau) \d_\zb\p'_c(0,0)\R \r] \NN\\
&=& {K_c\over2\pi^2}\l[ {1\over(v_c\tau-ix)^2}+{1\over(v_c\tau+ix)^2} \r]
\EEA
This, of course, is the quasi-uniform component of the compressibility. The
corresponding result for the magnetic susceptibility (still at $K_s=1$) is
\BE
\L S_z(x,\tau)S_z(0,0) \R = 
{1\over8\pi^2}\l[ {1\over(v_s\tau-ix)^2}+{1\over(v_s\tau+ix)^2} \r]
\EE
where $v_s$ is the spin velocity.

The compressibility also has a charge-density-wave component at $\pm 2k_F$,
determined by the correlations of the operator
\BEA
\Ocdw &=& \e^{-2ik_Fx} \sum_\s \psi_\s^\dg\psib_\s \NN\\
&=& {1\over2\pi}\e^{-2ik_Fx} \sum_\s \eta_\s\etab_\s \e^{-i\sk\p_\s} \NN\\
&=& {1\over2\pi}\e^{-2ik_Fx}\e^{i\sd\p_c} \l[
\eta_\up\etab_\up \e^{i\sd\p_s} + \eta_\dn\etab_\dn \e^{-i\sd\p_s}\r]\quad
\EEA
The CDW compressibility is then
\BEA
\L \Ocdw(x,\tau)\Ocdw^\dg(0,0)\R &=&
{1\over2\pi^2}\e^{-2ik_Fx} \L \e^{i\sd\p_c(z)}\e^{-i\sd\p_c(0)}\R 
\L \e^{i\sd\p_s(z)}\e^{-i\sd\p_s(0)}\R \NN\\
&=& {1\over2\pi^2}\e^{-2ik_Fx}
{1\over|v_c\tau-ix|^{K_c}}{1\over|v_s\tau-ix|}\quad
\EEA
Notice that the Klein factors disappear from the correlation function, as they
do in general from a product of the type ${\cal O}{\cal O}^\dg$.

\subsection{Spin or charge gap}

The interactions $\Ham_1$ and $\Ham_3$ do not lend themselves to an exact
solution like $\Ham_2$ and $\Ham_4$ do. However, they are subject to RG
arguments, at least at weak coupling.

\begin{figure}
\figEPS{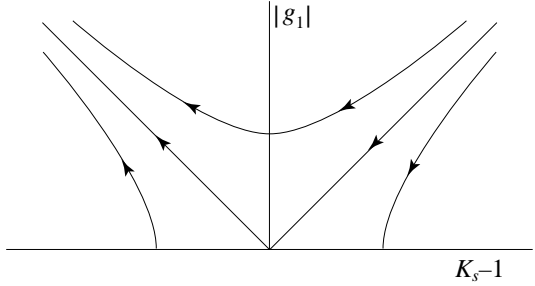}{7cm}
\caption{RG flow with backward scattering $g_1$ near the point
$K_s=1$.}\label{RGflow}
\end{figure}
Let us first consider the backward scattering term (\ref{bosint1}). Combined
with the spin boson Hamiltonian (for $K_s\ne1$), the 
perturbation defines the well-known sine-Gordon model, one of the most studied
one-dimensional field theories:\footnote{For the sake of argument, I have
neglected the velocity renormalization here ($v_s=v$). It would lead to an
additional renormalization of $g_1$, defined by $vg_1 = v_sg'_1$, which is
neglected at low coupling.}
\BE
\Ham_{\rm SG} =  \hf v \l[\Pi_s'^2 + (\d_x\p'_s)^2 \r] 
+ {vg_1\over2\pi^2}\cos\l(\sqrt{8\pi K_s}\p_s'\r)
\EE
(we used the rescaled boson $\p_s'$ in order to give the unperturbed part its
standard normalization).
If $K_s=1$, the cosine term is marginal, as can be checked from
Eq.~(\ref{dimvertex}): $\Delta=h+\hb=1+1=2$. One would naively conclude that it
is irrelevant if $K_s>1$, and relevant if $K_s<1$, thus leading to a gap in the
spin excitations. The story is a little more complicated by the existence at
$g_1=0$ not of a fixed {\it point}, but of a fixed {\it line}, parametrized by
$K_s$. It turns out that $K_s$ is subjected to RG flow away from that line and
one must instead consider the set of coupled RG equations
\BE\label{RGeq1}
\dd{K_s}{\ell} = -{1\over 2\pi^2} K_s^2 g_1^2 \qquad
\dd{g_1}{\ell} = -2 g_1(K_s-1)
\EE
first obtained by Kosterlitz and Thouless \cite{Kosterlitz73}.
These equations are demonstrated in Appendix A, from the short distance
product, or OPE, of the perturbations involved.
The RG flow is illustrated on Fig.~\ref{RGflow}. One checks that, at weak
coupling, the system flows towards strong coupling if $|g_1]>2\pi(K_s-1)$
($g_1$ is then said to be {\it marginally relevant}) while it flows to $g_1=0$
if $|g_1]<2\pi(K_s-1)$ ($g_1$ is then said to be {\it marginally irrelevant}).
Thus, we expect a spin gap for $g_1$ infinitesimal and $K_s<1$.\footnote{Note
that the sign of $g_1$ is of no consequence: it is the result of a gauge choice
through Klein factors. The arbitrariness in the sign of $g_1$ may be traced
back to its expression in terms of left and right fermions: changing the sign
of either
$\psi_\up$, $\psib_\up$, $\psi_\dn$ or
$\psib_\dn$ is of no consequence in the noninteracting case, but changes the
sign of $g_1$ and $g_3$, without affecting those of $g_2$ or $g_4$.}

This expectation is fullfilled exactly at the special value $K_s=\hf$. There,
the cosine term has scaling dimension 1 and is precisely what one would obtain
from bosonizing the Dirac mass term of Eq.~(\ref{DiracH}), except that the
corresponding fermion represents not electrons, but massive spinons. This
$K_s=1$ theory is called the Luther-Emery model \cite{Luther74} and
is a free theory for spinons, even though it is a complicated
interacting theory for electrons (and who knows what happens in the
charge sector at the same time\dots). 

The above analysis can be repeated for the Umklapp term (\ref{bosint3}), which
occurs only at half filling. One simply has to replace $g_1$ by $g_3$ and $K_s$
by $K_c$. Thus, a charge gap develops if $K_c<1$ (i.e. if $g_{2,c}$ is
repulsive) or if $|g_3]>2\pi(K_c-1)$. Again, free massive holons occur at
$K_c=\hf$.

\section{Non-Abelian bosonization}

\subsection{Symmetry currents}

The bosonization procedure described in Sect.~\ref{bosoS} is often called
{\it Abelian bosonization}, because a single compactified boson $\p$ has a
$U(1)$ symmetry $\p\to\p+a$ and the group $U(1)$ is Abelian (i.e. commutative).
Abelian bosonization may be used to describe the spin sector of a 1D electron
gas, but this description is not manifestly spin-rotation invariant. Indeed,
the chiral currents associated to the three spin components are easily shown
to be
\BEA\label{bosocurrents}
J^z &=& \hf \psi^\dg_\a (\s_3)_{\a\b}\psi_\b 
= {i\over\sd}\d_z\p_s \NN\\
J^x &=& \hf \psi^\dg_\a (\s_1)_{\a\b}\psi_\b 
= {i\over2\pi}\eta_\up\eta_\dn \sin\l(\sqrt{8\pi}\phi_s\r) \NN\\
J^y &=& \hf \psi^\dg_\a (\s_2)_{\a\b}\psi_\b 
= -{i\over2\pi}\eta_\up\eta_\dn \cos\l(\sqrt{8\pi}\phi_s\r)
\EEA
The $z$ component manifestly has a special role in this representation. While
the correlations $\L J^a(x)J^a(0)\R$ ($a=x,y,z$) all decay with the same power
law, they do not have the same normalization, a somewhat unsatisfactory feature.

Fortunately, there exists a bosonization procedure that is manifestly symmetric
under the spin rotation group SU(2), or under any Lie group for that matter.
This {\it non-Abelian bosonization} expresses a set of fermion fields in
terms of a matrix field belonging to a representation of a Lie algebra, instead
of one or more simple boson fields. The corresponding boson theory, the
Wess-Zumino-Witten (WZW) model, is far less familiar than the ordinary boson.
But in all situations where symmetry considerations (including symmetry
breaking) are important, non-Abelian bosonization is the method of
choice. It is of course unrealistic to give a rigorous and complete
introduction to the WZW model and non-Abelian bosonization within this
brief tutorial. Ref.~\cite{DMS97} may be consulted for an in-depth
discussion.

Let us first return to Abelian bosonization and insist on the role of the
U(1) current, before generalizing to larger symmetry groups. The currents
associated with the U(1) symmetry are given in Eq.~(\ref{U1currents}) for a
single species of fermions and
coincide with the chiral fermion density fluctuations. For the charge U(1)
symmetry, it is the total ($\up$+$\dn$) density fluctuation that matters,
and the charge currents are therefore
\BE
J_c = {i\over\sqrt{2\pi}}\d_z\p_c
\qquad \Jb_c = -{i\over\sqrt{2\pi}}\d_\zb\p_c
\EE
In terms of the $U(1)$
group element $g=\e^{i\p_c}$, these currents are
proportional to $g^{-1}\d_z g$ and $g^{-1}\d_\zb g$. The charge Hamiltonian can
then be expressed in terms of these currents, as in Eq.~(\ref{HU1}):
\BE
H ={\pi v_c\over2}\int\dx \l( J_c^2 + \Jb_c^2 \r)
\EE
This is enough to define the theory, provided that (i) the commutation relations
of the currents are specified:
\BE
[J_c(x),J_c(x')] = -{i\over\pi}\delta'(x-x') 
\EE
and (ii) the vertex operators are introduced, with their proper commutations
with the currents. Note: The above current commutation relation can be
demonstrated in the same way as we have recovered the fermion anticommutation
relations (\ref{anticomm}), from the current-current correlation function
\BE
\L J_c(z)J_c(z')\R = {1\over 8\pi^2}{1\over (z-z')^2}
\EE
obtained from Eq.~(\ref{phicorr}) by differentiating w.r.t. $z$ and $z'$.

Likewise, non-Abelian bosonization (for SU(2)) may be introduced starting with
the various components of the spin currents $J_a$ defined above in terms of two
electron fields $\psi_\s$, and expressed in terms of a group element $g$ as
\BE
J(z) = (\d_z g) g^{-1} = \hf\sum_a J_a\s_a \qquad 
\Jb(\zb) = g^{-1}(\d_\zb g)  = \hf\sum_a \Jb_a\s_a
\EE
The Hamiltonian of the SU(2) WZW model can
be expressed in terms of these SU(2) currents:
\BE
H = {2\pi v\over k+ 2}\sum_{a=1}^{3} \int \dx
\l[\no{J^a(x)J^a(x)} + \no{\Jb^a(x)\Jb^a(x)}\r]
\EE
where the integer $k$ is called the {\it level} of the WZW model, and where
the currents now obey the following commutation rules
\BEA\label{su2comm}
\com{J^a(x),J^b(x')}  &=&  -{ik\over2\pi}\delta_{ab}\delta'(x-x') + 
i \eps_{abe} J^e(x)\delta(x-x')\NN\\
\com{\Jb^a(x),\Jb^b(x')}  &=&  -{ik\over2\pi}\delta_{ab}\delta'(x-x') + 
i \eps_{abe} \Jb^e(x)\delta(x-x')\NN\\
\com{J^a(x),\Jb^b(x')} &=& 0 
\EEA
Once the currents are Fourier expanded as
\BE
J^a(z) = {1\over L}\sum_n J^a_n \e^{i2\pi nz/L}
\EE
The above commutators translate into the following set of commutation rules,
known as a {\it Kac-Moody algebra}:
\BEA
\com{J^a_m,J^b_n} &=& i\eps_{abe}J^e_{m+n} 
+ \hf k n\delta_{ab}\delta_{m+n,0}\NN\\
\com{\Jb^a_m,\Jb^b_n} &=& i\eps_{abe}\Jb^e_{m+n} +
\hf k n\delta_{ab}\delta_{m+n,0}
\EEA
States of the WZW model are created by applying the operators $J^a_n$ ($n<0$)
on the vacuum $|0\R$ and on a finite number of spin states.

The formulation of the WZW in terms of currents is not as fundamental as its
definition in the Lagrangian formalism, in terms of a field $g(x,t)$ taking
its values in SU(2) (or in another Lie group). The model must be such
that the currents are chiral, i.e. $\d_z\Jb=\d_\zb J=0$, and must
have conformal invariance. The needed action is
\BE\label{WZWaction}
S[g] = {k\over8\pi}\int \dr ^2x\; \tr(\d^\mu g^{-1}\d_\mu g)
-{ik\over 12\pi}\int_B \dr ^3x\; 
\eps_{\mu\nu\la}\tr \l( g^{-1}\d^\mu g\,g^{-1}\d^\nu g\,g^{-1}\d^\la g \r)\quad
\EE
where the second term is topological: it is integrated in a three-dimensional
manifold $B$ whose boundary is the two-dimensional space-time, but its value
depends only on the field configuration on the boundary of $B$. The {\it level}
$k$ must be an integer in order for the
WZW model to have conformal invariance and chiral currents. Working with the
action (\ref{WZWaction}) is somewhat unwieldy and almost never done in practice,
once key identities (Ward identities) have been derived, leading, among others,
to the commutators (\ref{su2comm}).

Bosonization cannot be done in terms of the currents alone. The group elements
$g$ occur in the representation of left-right products of fermions, as
specified by the {\it Witten formula}\cite{Witten84}:
\BEA\label{Witten}
\psi_\a\psib_\b^\dg &=& {1\over 2\pi}g_{\a\b} \e^{-i\sd\p_c} \NN\\
\psib_\a\psi_\b^\dg &=& {1\over 2\pi}(g^\dg)_{\a\b} \e^{i\sd\p_c}
\EEA
where the presence of the charge boson $\p_c$ is still necessary in order to
represent the charge component of the products. This formula leads to the
following representation of the charge and spin densities:
\BEA
2\pi \ntot(x) &=& J+\Jb 
+\l[\e^{-2ik_Fx} \e^{i\sd\p_c}\tr g + {\rm H.c.}\r] \NN\\
2\pi S^a_{\rm tot.}(x) &=& J^a+\Jb^a 
+\l[\e^{-2ik_Fx} \e^{i\sd\p_c}\tr(g\s_a) + {\rm H.c.}\r]
\EEA

The WZW model is a conformal field theory, with a set of well-identified
primary (or scaling) fields. It is completely soluble, in the sense that
the structure of its Hilbert space is completely known and correlation
functions obey linear differential equations. The central charge of the
level-$k$ SU(2) WZW model is
\BE
c = {3k\over k+2}
\EE
and it contains scaling fields of all spins from $s=0$ to $s=k/2$, with
conformal dimensions
\BE\label{WZWdim}
h = \hb = {s(s+1)\over k+2}
\EE
The spin part of the 1D electron gas with one band is described by the simplest
of all WZW models: SU(2) at level 1. The matrix field $g$ corresponds to $s=\hf$
and has conformal dimensions $(\frac14,\frac14)$. Its components may be
expressed in terms of the spin boson $\p_s$, if we follow Witten's formula and
the Abelian bosonization formulas:
\BE
g = \l( \BA{cc} \e^{-i\sd\p_s} & \e^{-i\sd\dual_s} \\
 \e^{i\sd\p_s} & \e^{i\sd\dual_s} \EA\r)
\EE
The equivalence of the $k=1$ SU(2) WZW model with the spin boson theory is
possible because the $k=1$ model has central charge unity. Such a
correspondence between a WZW model and simple bosons is possible only if $c$
is an integer. 

Other WZW models describe more exotic critical systems. The biquadratic
spin-1 chain, with Hamiltonian
\BE
H = \sum_n {\bf S}_n\cdot{\bf S}_{n+1} + \b({\bf S}_n\cdot{\bf S}_{n+1})^2
\EE
is critical at $\b=1$ and $\b=-1$. At $\b=-1$, the low-energy limit is
described by the $k=2$ SU(2) WZW model, which has central charge $c=\frac32$.
This model may be expressed in terms of three Majorana
fermions~\cite{Zamolodchikov86,Tsvelik90}. At $\b=+1$, the low-energy limit is
described by the $k=1$ SU(3) WZW model. A recent attempt at extending Zhang's
SO(5) theory of antiferromagnetism and d-wave superconductivity to
one-dimensional systems is formulated in terms of WZW models~\cite{Shelton98}.

\subsection{Application to the perturbed Tomonaga-Luttinger model}
\label{NabosTLSS}

The noninteracting part of the Tomonaga-Luttinger Hamiltonian may be expressed
as follows in terms of charge and spin currents:
\BE\label{TL2}
\Ham_0 = {\pi v\over2} \l[ J_c^2 + \Jb_c^2 \r]
+  {2\pi v\over 3} \l[ \Jv^2 + \Jbv^2 \r]
\EE
It is the sum of a free charge boson and of a $k=1$ SU(2) WZW model. The
various interactions bosonized in Sect.~\ref{bosointSS} have the following
expressions in terms of currents:
\BEA
\Ham_1 &=& -2vg_1 (J_x\Jb_x+J_y\Jb_y) \NN\\
\Ham^c_2 &=& vg_{2,c} J_c\Jb_c \NN\\
\Ham^c_4 &=& \hff vg_{4,c} [J_c^2+\Jb_c^2] \NN\\
\Ham^s_2 &=& 4vg_{2,s} J_z\Jb_z \NN\\
\Ham^s_4 &=& {2v\over3}g_{4,s}\l[ \Jv^2 + \Jbv^2 \r]\NN\\
\Ham_3 &=& {vg_3\over2\pi^2} \cos(\sqrt{8\pi}\p_c) 
\EEA
This correspondence may be established with the help of
Eq.~(\ref{bosocurrents}). It is then manifest that $g_{4,c}$ and $g_{4,s}$ only
renormalize the velocity. Moreover, if $g_1=-2g_{2,s}$, $\Ham_1$ and $\Ham^s_2$
combine into a single, rotation-invariant interaction:
\BE
\Ham_1 + \Ham^s_2 = 4vg_{2,s} \Jv\cdot\Jbv \qquad (g_1=-2g_{2,s})
\EE

\subsubsection*{The spin sector}
Let us restrict ourselves to this isotropic case, at half-filling, so that
$g_3\ne0$. As we have seen in Fig.~\ref{RGflow}, the RG flow will bring $g_3$ to
strong coupling if $|g_3|>-2g_{2,c}$ and the charge sector will have a gap. At
the same time, the coupling $g_1$ flows along the separatrix of
Fig.~\ref{RGflow}, towards the fixed point $g_1=0, K_s=1$ (a detailed RG
analysis, following the method described in Appendix A, shows that the
separatrix is precisely determined by the condition of rotation invariance
$g_1=-2g_{2,s}$). Thus, the low-energy limit of the half-filled perturbed TL
model is the $k=1$ WZW model, provided
$|g_3|>-2g_{2,c}$ and 
$g_1=-2g_{2,s}$.

The appearance of a gap in the charge sector has another consequence, that of
changing the scaling dimensions of some operators. Specifically, let us
consider the $2k_F$ component of the spin density, which becomes the staggered
magnetization at  half-filling:
\BEA\label{stagmag}
S_a^{2k_F} &=& \hff \psib^\dg_i(\s_a)_{ij}\psi_j + {\rm H.c.} \NN\\
&=& -{1\over4\pi}g_{ji}(\s_a)_{ij}\e^{i\sd\p_c} + {\rm H.c.} \NN\\
&=& -{1\over4\pi}\tr(\s_a g)\e^{i\sd\p_c} + {\rm H.c.}
\EEA
where we have applied Witten's formula (\ref{Witten}).
The cosine Umklapp term gives $\p_c$ a nonzero expectation value, such
that $\L\cos\sk\p_c\R\ne0$ and $\L\sin\sk\p_c\R=0$. Neglecting fluctuations
around that expectation value, one may write the staggered magnetization
density as
\BE\label{stagmag2}
S_a^{2k_F} = {\rm const.}\l(\tr(\s_a g) + {\rm H.c.}\r)
\EE
According to Eq.~(\ref{WZWdim}) with $s=\hf$, this operator has scaling
dimension $\Delta=h+\hb=\hf$, although it had $\Delta=1$ initially (before the
onset of the charge gap). Essentially, the charge contribution to the
fluctuations, i.e. the operator $\e^{\pm i\sd\p_c}$, also of scaling dimension
$\Delta=\hf$, is frozen; this makes the operator $S_a^{2k_F}$ more relevant. The
staggered magnetization correlation is therefore decreasing like
$1/r$, instead of the $1/r^2$ decay of uniform correlations:
\BE\label{stagmagcorr}
\L S_a^{2k_F}(x,t)S_a^{2k_F}(0,0)\R = {{\rm const.}\over |x^2-v^2t^2|^{1/2}}
\EE

\subsubsection*{Other Perturbations}

One may add various relevant or marginally relevant perturbations to the
model (\ref{TL2}):
\BI
\item An Ising-like anisotropy would contribute a term of the form
$\Ham_{\rm Ising}= \la J_z\Jb_z$. This term, a correction to $g_{2,s}$ is
marginally relevant if $\la>0$: it brings the flow above the separatrix of
Fig.~\ref{RGflow}. On the other hand, it is marginally irrelevant if
$\la<0$, and brings the system towards an XY fixed point.
\item An explicit dimerization, maybe in the form of a staggered hopping term in
the underlying Hubbard model. This perturbation would be proportional to the
CDW (i.e. $2k_F$) component of the electron density. In terms of the WZW field
$g$, this may be obtained by replacing the Pauli matrices of
Eq.~(\ref{stagmag}) by the unit matrix. At half-filling, the freezing of
$\cos\sk\p_c$ would leave the single scaling field $\tr g$ in the perturbation:
\BE
\Ham_{\rm dim.} = \la\tr g
\EE
This term has scaling dimension $\Delta=\hf$ and would lead, according to
Eq.~(\ref{gap}), to a spin gap of order $m\sim \la^{2/3}$.
\item A uniform external magnetic field $B$. This couples to the uniform
magnetization density $J_z+\Jb_z$, producing a perturbation with
nonzero conformal spin. 
It breaks rotation invariance and is best treated within Abelian bosonization.
In terms of the spin boson $\p_s$, the perturbation is $\la\d_x\p_s$, where
$\la\propto B$. The spin Hamiltonian is then
\BE
\Ham = \hf v \l[\Pi_s^2 + (\d_x\p_s)^2 \r] + \la\d_x\p_s
\EE
(we have supposed $K_s=1$, i.e., rotation invariance in the absence of external
field). This perturbation can be eliminated from the Hamiltonian by redefining 
\BE
\p_s \to \p_s - {\la\over v} x
\EE
In terms of the original electrons, this is almost equivalent to changing the
value of $k_F$ (almost, because it should have no effect on the charge sector).
Basically, the uniform magnetic field introduces some incommensurablility in
the spin part of the system.
\EI

\section{Other applications of bosonization}

\subsection{The spin-$\hf$ Heisenberg chain}

The spin-$\hf$ Heisenberg model in one dimension may also be treated by
bosonization in the low-energy limit. The microscopic Hamiltonian is
\BE
H = \sum_n \l\{
J(S^x_n S^x_{n+1} + S^y_n S^y_{n+1}) + J_z S^z_n S^z_{n+1} \r\}
\EE
This Hamiltonian may be mapped to a spinless fermion problem via the
Jordan-Wigner transformation:
\BE
S_n^+ = c^\dg_n\exp\l(i\pi\sum_m^{n-1}c_m^\dg c_m\r) \qquad
S_n^z  = c_n^\dg c_n-\hf
\EE
where the operators $c_n$ and $c_m^\dg$ obey standard anticommutation relations.
In terms of these operators, the Heisenberg Hamiltonian is
\BE
H = \sum_n \l\{-{J\over2}(c_n^\dg c_{n+1}+{\rm H.c.}) 
+ J_z(c_n^\dg c_n-\hf) (c_{n+1}^\dg c_{n+1}-\hf) \r\}
\EE
The fermion number $\sum_n c_n^\dg c_n$ is related to the total magnetization
$M$, in such a way that half-filling corresponds to $M=0$, the
appropriate condition in the absence of external magnetic field.
The continuum limit may then be taken and the bosonization
procedure applied. The spinless fermion
is then endowed with interactions proportional to $J_z$. Two of these
interaction terms are of the $g_2$ and
$g_4$ type and renormalize the boson field $\p$ and the
velocity $v$. From comparing with known exact exponents, one derives the
following relation between the radius $R$ of the boson and the bare interaction
strength $J_z$:
\BE
R = {1\over\sqrt 2\pi}\l\{ 1 - 
{1\over\pi}\cos^{-1}\l({J_z\over J}\r)\r\}^{1/2}
\EE
Various correlation functions may be calculated, once the correspondence
between spin components and the boson field is known~\cite{Affleck89}:
\BEA
S^z(x) &=& {1\over2\pi R}\d_x\p + (-1)^x {\rm cst.} \e^{i\p/R} \NN\\
S^-(x) &=& \e^{-2\pi i R\dual} + (-1)^x {\rm cst.}\l\{
\e^{i[\p/R-2\pi R\dual]}+\e^{i[-\p/R-2\pi R\dual]} \r\}\quad
\EEA
For instance,
\BEA
\L S^z(x,t) S^z(0,0)\R &=& {1\over 16\pi^3 R} \l(
\L \d_z\p(x,t)\d_z\p(0,0)\R + \L \d_\zb\p(x,t)\d_\zb\p(0,0)\R\r) 
\NN\\ && \qquad
+ {\rm cst.}(-1)^x \L \e^{i\p(x,t)/R}\e^{i\p(0,0)/R} \R \NN\\
&=& {1\over 16\pi^3 R} \l[{1\over(x-vt)^2}+{1\over(x+vt)^2}\r] 
+ {\rm cst.}(-1)^x {1\over|x^2+v^2t^2|^{1/4\pi R^2}} 
\EEA
Note that the staggered part has a nonuniversal exponent, which depends on
$J_z/J$. At the isotropic point $J_z=J$, $R=1/\sd$ and this exponent is
$\hf$, exactly like the non-Abelian bosonization prediction (\ref{stagmagcorr}).

\subsection{Edge States in Quantum Hall Systems}

\begin{figure}
\figEPS{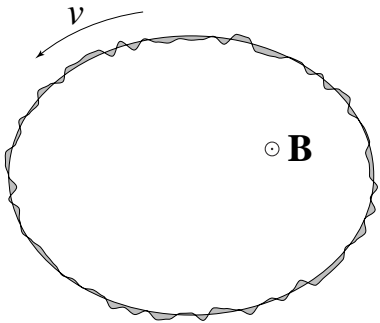}{5cm}
\caption{Fluctuation of the edge of a quantum Hall system.
}
\label{edge}
\end{figure}

The two-dimensional electron gas (2DEG) in a magnetic field has been the object
of intense theoretical investigation since the discovery of the quantum
Hall effect (integral and fractional) in the early 1980's.
While the integer Hall effect may be understood (in the bulk) on the basis of
weakly correlated electrons, the fractional quantum Hall effect is a strongly
correlated problem. Quantum Hall plateaus correspond to gapped states, but
electric conduction nevertheless occurs through electrons at the edge of the 2D
gas, the so-called {\it edge states}. One must imagine a 2DEG immersed in a
magnetic field $B$ and further confined to a finite area by an in-plane electic
field $E$ (in, say, a radial configuration). The field causes a persistent
current along the edge of the gas:
\BE
{\bf j} = \s_{xy}{\bf\hat z}\wedge{\bf E}
\EE
where $\s_{xy}$ is the Hall conductance, equal to $\nu e^2$, where $\nu$ is the
filling fraction. The edge electrons drift in one direction at the velocity
$v=cE/B$. We will adopt a hydrodynamic description of the edge, following
Ref.~\cite{Wen92}. Let $x$ is the coordinate along the edge and $h(x)$
the transverse displacement of the edge with respect to the ground state shape
of the gas (cf. Fig.~\ref{edge}). If $n=\nu/2\pi\ell_B^2$ is the electron
density ($\ell_B$ is the magnetic length), then the linear electron density in
an edge excitation is
$J(x)=nh(x)$. The energy of the excitation is the electrostatic energy
associated with the edge deformation:
\BE
H = \int\dx \hf e^2 hJ E = \pi\nu v\int\dx J^2
\EE

The field $J$ being from the start treated as a collective mode,
bosonization is natural. The main difference with the 1D electron gas is
that only one half (left of right) of the boson is necessary, since the edge
excitations only propagate clockwise or counterclockwise, depending on the
field direction. We therefore assert
\BE
J = {1\over\sqrt\pi}\d_x\phi = -{i\over\sqrt\pi}\d_z\phi
\EE
and the Hamiltonian density for edge excitations is
\BE
\Ham = -v\nu(\d_z\phi)^2
\EE
The filling fraction $\nu$ plays the role of a field renormalization $K=1/\nu$.
The edge electron must be represented by a field $\psi(x)$ such that
\BE
[J(x),\psi(x')] = -\delta(x-x')\psi(x)
\EE
From the short-distance product of $J$ with $\e^{-i\a\phi}$, one finds that
the correct representation for $\psi$ is
\BE
\psi(x) = {1\over\sd} \e^{i\sk\phi/\nu}
\EE
From this we derive the one-electron Green function
\BE\label{QHEGreen}
G(x,t) = {1\over2\pi} {1\over(x-vt)^{1/\nu}}
\EE
Note that this coincides with a free-electron propagator in the case $\nu=1$
only, i.e., when the first Landau level is exactly filled. Even in the integer
effect, the edge system is a Luttinger liquid if $m=1/\nu>1$, albeit a {\it
chiral} Luttinger liquid. In momentum space, the Green function (\ref{QHEGreen})
is
\BE
G(k,\om) \propto {(\om+vk)^{m-1}\over \om-vk}
\EE
and the momentum-integrated density of states near the Fermi level is
\BE
N(\om) \propto |\om|^{m-1}
\EE

\subsection{And more\dots}

This short tutorial has concentrated on basic Luttinger liquid physics. Time
and space are lacking to introduce the many physical
systems that have been treated with bosonization. Here we will just indicate a
few useful references.

The bosonization formalism may also be used to treat disorder. The problem of
an isolated impurity in a Luttinger liquid is treated in Refs.~\cite{Gogolin98}
and \cite{Schulz98}, following Kane \& Fisher's original work~\cite{Kane92}.
The more complicated problem of Anderson localization, resulting from the
coherent scattering off many impurities, is also described summarily in
Ref.~\cite{Schulz98}. Formally related to impurity scattering is the question
of electron-phonon interaction in a Luttinger liquid, a review of which may be
found in Ref.~\cite{Gogolin98}.

The Kondo problem, or the strong interaction of magnetic impurities with
conduction electrons, has also been treated using bosonization, and a strong
dose of conformal field theory, by Affleck and Ludwig. Ref.~\cite{Affleck95} is
a comprehensive review of the progress accomplished with the help of conformal
methods. Most interesting in this application is the three-dimensional nature
of the initial problem, before it is mapped to a one dimension.

Spin chains of spin $s>\hf$ can be constructed from $2s$ fermion species with a
Hund-type coupling. This is the approach followed in Ref.~\cite{Affleck87},
also reviewed in Ref.~\cite{Affleck89}. The spin-1 chain has been treated as a
perturbed $k=2$ SU(2) WZW model in Ref.~\cite{Tsvelik90}.

Coupled spin chains, or spin ladders, have been studied in Abelian bosonization
and are reviewed in Ref.~\cite{Schulz98}. A non-Abelian bosonization study of
two coupled spin chains is performed in Ref.~\cite{Allen97}. Doped spin ladders
are coupled Luttinger liquids. Bosonization studies may be found in
Refs.~\cite{Fabrizio93} and \cite{Schulz96}. Ref.~\cite{Shelton98} looks at
this problem from the perspective of SO(5) symmetry. The latter two references
treat the question of d-wave fluctuations in a doped ladder.

The reader is referred to Ref.~\cite{Schulz98,Gogolin98} for additional
references (we apologize for our poor bibliography). More applications of
bosonization to condensed matter can be found in Ref.~\cite{Gogolin98}. Other
important review articles on the subject include Refs.~\cite{Schulz98} and
\cite{Affleck89}. A historical perspective is given in the collection of
preprints \cite{Stone94}.

\section{Conclusion}

Let us conclude by highlighting the principal merits and
limitations of bosonization:
\BI
\item Bosonization is a {\it nonperturbative} method. The Tomonaga-Luttinger
(TL) model, a continuum theory of {\it interacting} fermions, can be translated
into a theory of noninteracting bosons and solved exactly.
\item Bosonization is a method for translating a fermionic theory into a
bosonic theory, and eventually retranslating part of the latter into a new
fermionic language (cf. the Luther-Emery model or the use of Majorana
fermions). This translation process is exact in the continuum limit, but does
not warrant an exact solution of the model, except in a few exceptional cases
(e.g. the TL model). For the rest, one must rely on renormalization-group
analyses, which generally complement bosonization.
\item The bosonized theory may have decoupled components, like spin and charge,
which are not manifest in fermionic language. Thus, spin-charge separation is an
exact prediction of bosonization (even beyond the TL model). When one of the
components (spin or charge) becomes massive, this favors the emergence of
physical operators with new scaling dimensions, like the staggered
magnetization of Eq.~(\ref{stagmag2}).
\item The contact with microscopic model is not obvious, since an infinite
number of irrelevant parameters stand in the way. But bosonization leads to
universal or quasi-universal predictions, and one can argue that such
predictions are preferable to the exact solution of a particular microscopic
Hamiltonian of uncertain relevance.
\item In the rare cases where a microscopic Hamiltonian has an exact solution
(e.g. the 1D Hubbard model), the solution is so complex that dynamical
quantities cannot be explicitly calculated. Bosonization can then be used as a
complementary approach. Exact values of thermodynamic quantities of the
microscopic theory may be used to fix the parameters (e.g. $v_c$, $v_s$, $K_c$,
$K_s$) of the boson field theory, and the latter may be used to calculate
dynamic quantities (e.g. the spectral weight) or asymptotics of correlation
functions\cite{Schulz90,Kawakami92}.
\item The low-energy limit corresponds to a few regions in momentum-space where
bosonization has something to say: the vicinity of wavevectors that are small
multiples of $k_F$.
\item Finally, bosonization is limited to one-dimensional systems, despite
attempts at generalizing it to higher dimensions
\cite{Haldane92,Kwon95}.
\EI

\section*{Acknowledgments}
The author expresses his gratitude to the Centre de Recherches Math\'ematiques
for making this workshop possible, and to the organizers for inviting him.
Support from NSERC (Canada) and FCAR (Qu\'ebec) is gratefully acknowledged.

\appendix
\section{RG flow and Operator Product Expansion}

\def\Phit{\tilde\Phi}
\def\xv{{\bf x}}
\def\yv{{\bf y}}
\def\Xv{{\bf X}}
\def\wb{{\bar w}}

In this appendix we show how to derive the approximate RG trajectory around
a fixed-point from the Operator Product Expansion (OPE) of the marginal
perturbations involved. Consider a fixed-point action $S_0$ describing a
conformal field theory, and a set of perturbations with coupling constants
$g_i$:
\BE
S[\Phi] = S_0[\Phi] + \sum_i g_i \int \dr x\dr\tau\; O_i(x,\tau)
\EE
Here $\Phi$ denotes a field or a collection of fields that defines the theory,
in the sense of the action and of the path integral measure. The local operators
$O_i$ are in principle functions of $\Phi$.
The RG procedure may be performed in momentum space or in
direct space. It is more convenient here to used the direct space approach and to
divide space-time into tiles of side $L$. A space-time point may then be
parametrized as
\BE
\xv = \Xv + \yv
\EE
where $\xv$ is a space-time point, $\Xv$ is the space-time coordinate of a tile
and $\yv$ a coordinate within a tile. The RG step is performed by tracing over
degrees of freedom labeled by $\yv$, leaving us with ``block-spin'' degrees of
freedom $\Phit(\Xv)$. This is done approximately be expanding the
perturbations to second order in the partition function, in order to construct
an effective action for the ``block spin'' variables:
\BE\label{Zexpansion}
\exp-S_{\rm eff.}[\Phit] \approx \e^{-S_0[\Phit]}\l\{
1 - \sum_i g_i \int\dr\xv \L O_i(\xv)\R_> 
+ \hf\sum_{i,j} g_i g_j\int\dr\xv\dr\xv' \L O_i(\xv)O_j(\xv')\R_> +\cdots\r\}
\EE
where the expectations values are taken over the fast modes, i.e., the modes
that vary within a tile, in the fixed-point theory. Notice that the fixed-point
action $S_0$ is the same for $\Phi$ as for the block-spin field $\Phit$, precisely
because $S_0$ represents a fixed point.

Consider first the integral
\BE
\int\dr\xv \L O_i(\xv)\R_> 
= \sum_\Xv \int_{\rm tile}\dr\yv \L O_i(\Xv + \yv)\R_>
\EE
The integration over the tile of side $L$ may be inferred from dimensional
analysis: if $(h_i,\hb_i)$ are the conformal dimensions of the operator $O_i$
and $\Delta_i=h_i+\hb_i$ is the corresponding scaling dimension,\footnote{For
simplicity, we will deal with operators of zero conformal spin only, i.e.,
we suppose that $h_i=\hb_i$. Operators with nonzero conformal spin
naively do not contribute: they have zero expectation value.} then this
integral must be $L^{2-\Delta_i}$ (there is no other quantity with the correct
scaling dimension, since the correlation length of the fixed point theory is
infinite), times some effective $\Xv$ dependence, so that the first term of
Eq.~(\ref{Zexpansion}) may be written as
\BE
-\sum_i\sum_\Xv L^{2-\Delta_i} O_i(\Xv)
\EE
where the field $O_i(\Xv)$ denotes this $\Xv$ dependence.

The second term of Eq.~(\ref{Zexpansion}) may be treated in the same way, once
the product of operators has been reduced to a sum of single operators by the
OPE:
\BE
O_i(\xv)O_j(\xv') = \sum_k C_{ij}^k O_k(\xv'){1\over
|\xv-\xv'|^{\Delta_i+\Delta_j-\Delta_k}}
\EE
It is assumed here that the set of operators $O_i$ forms a ``closed algebra''
under this short-distance expansion. Only the most divergent terms really matter.
Let us substitute this OPE in the second term of Eq.~(\ref{Zexpansion}) and
evaluate the integrated expectation value like we did for the first term: we find
$L^{2-\Delta_k}$, times $O(\Xv)$, times the integral over the relative coordinate
$\xv-\xv'$, which scales like $L^{2-\Delta_i-\Delta_j+\Delta_k}$, or like $\ln L$
if $2=\Delta_i+\Delta_j-\Delta_k$. The effective action, after this RG step, is
thus given by
\BEA
\exp-S_{\rm eff.}[\Phit] &\approx& \e^{-S_0[\Phit]}\Bigg\{
1 - \sum_k g_k \int\dr\Xv\; L^{2-\Delta_k}O_k(\Xv) \NN\\
&& + \hf\sum_{k}\sum_{i,j} g_i g_j C_{ij}^k 
\left\lbrace
\matrix{
{L^{2-\Delta_i-\Delta_j+\Delta_k}} \cr \ln L
}
\right\rbrace
L^{2-\Delta_k} \int\dr\Xv\; O_k(\Xv) +\cdots\Bigg\}
\EEA
(a sum over tiles is equivalent to an integral over $\Xv$, since $L$ is the new
`lattice spacing'). After reinstalling the perturbations in the exponential, one
may define a renormalized coupling $g'_k$ as
\BE
g'_k = g_k L^{2-\Delta_k} - \hf\sum_{i,j} C_{ij}^k g_i g_j
\left\lbrace
\matrix{
{L^{2-\Delta_i-\Delta_j+\Delta_k}} \cr \ln L
}
\right\rbrace
L^{2-\Delta_k}
\EE

This RG step transformation must be made infinitesimal in order to be translated
into an RG flow equation. To this end, we set $L$ very close to unity, in units
of lattice spacing, a somewhat formal procedure. In the case of a
nonmarginal interaction ($\Delta_k\ne2$) the first term dominates in a
perturbative sense and, after setting $L=\e^\ell$ with $\ell$ small, we find
\BE
g'_k \approx g_k\l[1+(2-\Delta_k)\ell\r] ~~\rightarrow~~
{\dr g_k\over\dr\ell} = (2-\Delta_k)g_k
\EE
as expected. A more interesting case is that of a set of marginal interactions
($\Delta_k=2$), where
\BE\label{betafunc}
g'_k = g_k - \hf\sum_{i,j} C_{ij}^k g_i g_j \ln L ~~\rightarrow~~
{\dr g_k\over\dr\ell} = - \hf\sum_{i,j} C_{ij}^k g_i g_j
\EE

Let us apply this last equation to the spin sector of the 1D electron gas
subjected to the interactions $g_1$ and $g_2$, in order to recover the flow
equations (\ref{RGeq1}). Let us first note that, when
expanded to first order in $g_{2,s}$ from Eq.~(\ref{Ketv}), those flow equations
become
\BE\label{RGeq2}
\dd{g_1}{\ell} = {2\over\pi} g_1 g_2 \qquad
\dd{g_2}{\ell} = {1\over 2\pi} g_1^2 
\EE
(we dropped the `$s$' index from now on: it is understood that we deal with the
spin sector). The corresponding operators are
\BE
O_1 = {1\over 2\pi^2}\cos(\sqrt{8\pi}\p) \qquad
O_2 = {2\over\pi} \d\p\db\p
\EE
The corresponding OPEs are obtained by applying Wick's theorem. Their most
singular terms are
\BEA
O_1(z,\zb)O_1(w,\wb) &\sim& 
{1\over16\pi^4}{1\over (z-w)^2(\zb-\wb)^2}
-{1\over2\pi^3}\l[{1\over (z-w)^2}(\db\p)^2+{1\over (\zb-\wb)^2}(\d\p)^2\r]\NN\\
&&\qquad-{1\over\pi^3}{1\over (z-w)(\zb-\wb)}\d\p\db\p \NN\\
O_2(z,\zb)O_2(w,\wb) &\sim& 
{1\over4\pi^4}{1\over (z-w)^2(\zb-\wb)^2}
-{1\over\pi^3}\l[{1\over (z-w)^2}(\db\p)^2+{1\over (\zb-\wb)^2}(\d\p)^2\r] \NN\\
O_2(z,\zb)O_1(w,\wb) &\sim&
-{1\over2\pi^4} {1\over (z-w)(\zb-\wb)}\cos(\sqrt{8\pi}\p)
\EEA
This last OPE is derived from the relation
\BE
\d\p(z)\e^{i\a\p(w,\wb)} \sim -{i\a\over4\pi}{1\over z-w}\e^{i\a\p(w,\wb)}
\EE
which is obtained by applying Wick's theorem to the series expansion of
$\e^{i\a\p}$ (see, e.g., Ref.~\cite{DMS97}). We may rewrite these OPEs as
\BEA
O_1(z,\zb)O_1(w,\wb) &\sim& 
{\rm const.}
-{1\over2\pi^2} {1\over (z-w)(\zb-\wb)}O_2(w,\wb)
+{\rm conf.~spin}
\NN\\
O_2(z,\zb)O_2(w,\wb) &\sim& 
{\rm const.}
+{\rm conf.~spin}\NN\\
O_2(z,\zb)O_1(w,\wb) &\sim &
-{2\over\pi^2} {1\over (z-w)(\zb-\wb)}O_1(w,\wb)
\EEA
where `conf. spin' denotes terms with conformal spin, which do not affect
the beta functions as calculated above. From Eq.~(\ref{betafunc}), we may then
write the flow equations~:
\BE
\dd{g_1}{\ell} = {1\over\pi^2} g_1 g_2 \qquad
\dd{g_2}{\ell} = {1\over 4\pi^2} g_1^2 
\EE
These differ from Eq.~(\ref{RGeq2}) only by a rescaling of the flow variable
$\ell$ by $2\pi$, caused by our crude treatment of kinematic integrals. A
corrected version of Eq.~(\ref{betafunc}) would then be
\BE\label{betafunc2}
{\dr g_k\over\dr\ell} = -\pi\sum_{i,j} C_{ij}^k g_i g_j
\EE


\end{document}